\documentclass{elsarticle}

\usepackage[reviewcopy]{adndt}
\usepackage{longtable}

\usepackage{amsmath}

\usepackage{amssymb}

\biboptions{square,sort&compress}
\bibpunct[]{[}{]}{,}{n}{}{;}
\begin{document}

\begin{frontmatter}

\journal{Atomic Data and Nuclear Data Tables}


\title{Discovery of Rubidium, Strontium, Molybdenum, and Rhodium Isotopes}

\author{A. M. Parker}
\author{M. Thoennessen\corref{cor1}}\ead{thoennessen@nscl.msu.edu}

 \cortext[cor1]{Corresponding author.}

 \address{National Superconducting Cyclotron Laboratory and \\ Department of Physics and Astronomy, Michigan State University, \\ East Lansing, MI 48824, USA}

\begin{abstract}
Currently, thirty-one rubidium, thirty-five strontium, thirty-five molybdenum and thirty-eight rhodium isotopes have been observed and the discovery of these isotopes is discussed here. For each isotope a brief synopsis of the first refereed publication, including the production and identification method, is presented.
\end{abstract}

\end{frontmatter}





\newpage
\tableofcontents
\listofDtables

\vskip5pc

\section{Introduction}\label{s:intro}

The discovery of rubidium, strontium, molybdenum, and rhodium isotopes is discussed as part of the series summarizing the discovery of isotopes, beginning with the cerium isotopes in 2009 \cite{2009Gin01}. Guidelines for assigning credit for discovery are (1) clear identification, either through decay-curves and relationships to other known isotopes, particle or $\gamma$-ray spectra, or unique mass and Z-identification, and (2) publication of the discovery in a refereed journal. The authors and year of the first publication, the laboratory where the isotopes were produced as well as the production and identification methods are discussed. When appropriate, references to conference proceedings, internal reports, and theses are included. When a discovery includes a half-life measurement the measured value is compared to the currently adopted value taken from the NUBASE evaluation \cite{2003Aud01} which is based on the ENSDF database \cite{2008ENS01}. In cases where the reported half-life differed significantly from the adopted half-life (up to approximately a factor of two), we searched the subsequent literature for indications that the measurement was erroneous. If that was not the case we credited the authors with the discovery in spite of the inaccurate half-life.

The first criterion is not clear cut and in many instances debatable. Within the scope of the present project it is not possible to scrutinize each paper for the accuracy of the experimental data as is done for the discovery of elements \cite{1991IUP01}. In some cases an initial tentative assignment is not specifically confirmed in later papers and the first assignment is tacitly accepted by the community. The readers are encouraged to contact the authors if they disagree with an assignment because they are aware of an earlier paper or if they found evidence that the data of the chosen paper were incorrect. Measurements of half-lives of a given element without mass identification are not accepted. This affects mostly isotopes first observed in fission where decay curves of chemically separated elements were measured without the capability to determine their mass. Also the four-parameter measurements (see, for example, Ref. \cite{1970Joh01}) were, in general, not considered because the mass identification was only $\pm$1 mass unit.

The second criterion affects especially the isotopes studied within the Manhattan Project. Although an overview of the results was published in 1946 \cite{1946TPP01}, most of the papers were only published in the Plutonium Project Records of the Manhattan Project Technical Series, Vol. 9A, ''Radiochemistry and the Fission Products,'' in three books by Wiley in 1951 \cite{1951Cor01}. We considered this first unclassified publication to be equivalent to a refereed paper.
Good examples why publications in conference proceedings should not be considered are $^{118}$Tc and $^{120}$Ru which had been reported as being discovered in a conference proceeding \cite{1996Cza01} but not in the subsequent refereed publication \cite{1997Ber01}. For almost simultaneous reports of discovery credit is given to the paper with the earlier submission date. In these instances the later papers are also mentioned.

The initial literature search was performed using the databases ENSDF \cite{2008ENS01} and NSR \cite{2008NSR01} of the National Nuclear Data Center at Brookhaven National Laboratory. These databases are complete and reliable back to the early 1960's. For earlier references, several editions of the Table of Isotopes were used \cite{1940Liv01,1944Sea01,1948Sea01,1953Hol02,1958Str01,1967Led01}. A good reference for the discovery of the stable isotopes was the second edition of Aston's book ``Mass Spectra and Isotopes'' \cite{1942Ast01}.

\section{Discovery of $^{73-103}$Rb}

Thirty-one rubidium isotopes from A = 73--103 have been discovered so far; these include 2 stable, 12 neutron-deficient and 17 neutron-rich isotopes. According to the HFB-14 model \cite{2007Gor01}, $^{118}$Rb should be the last odd-odd particle stable neutron-rich nucleus while the odd-even particle stable neutron-rich nuclei should continue through $^{127}$Rb. The proton dripline has been reached at $^{74}$Rb because of the non-observance of $^{73}$Rb \cite{1991Moh01,1992Yen01}. A proton-unbound excited state has been observed in $^{73}$Rb \cite{1993Bat01}. $^{70}$Rb, $^{71}$Rb, and $^{72}$Rb could still have half-lives longer than 10$^{-9}$~ns \cite{2004Tho01}. About 23 isotopes have yet to be discovered corresponding to 43\% of all possible rubidium isotopes.

\begin{figure}
	\centering
	\includegraphics[scale=.5]{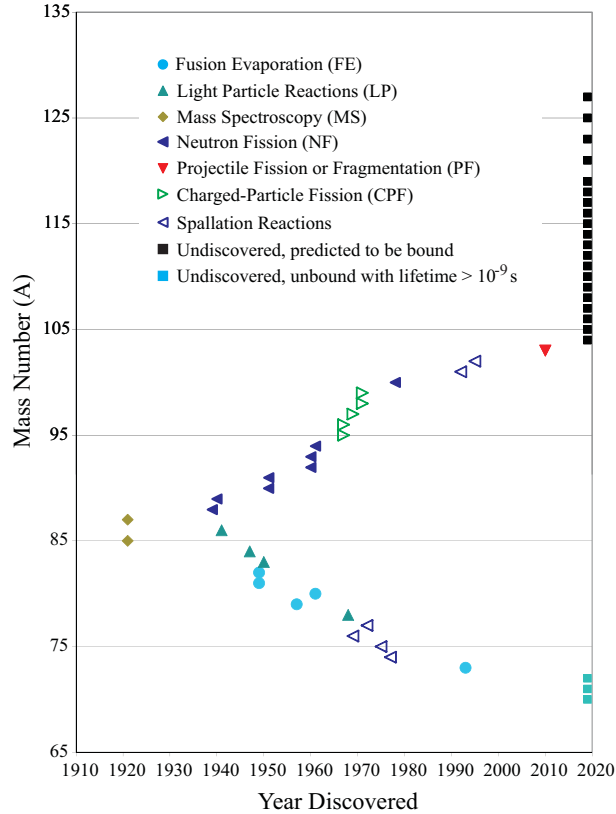}
	\caption{Rubidium isotopes as a function of time when they were discovered. The different production methods are indicated. The solid black squares on the right hand side of the plot are isotopes predicted to be bound by the HFB-14 model. On the proton-rich side the light blue squares correspond to unbound isotopes predicted to have lifetimes larger than $\sim 10^{-9}$~s.}
\label{f:year-rubidium}
\end{figure}

Figure \ref{f:year-rubidium} summarizes the year of first discovery for all strontium isotopes identified by the method of discovery. The range of isotopes predicted to exist is indicated on the right side of the figure. The radioactive strontium isotopes were produced using fusion evaporation reactions (FE), light-particle reactions (LP), neutron induced fission (NF), charged-particle induced fission (CPF), spallation reactions (SP), and projectile fragmentation or fission (PF). The stable isotopes were identified using mass spectroscopy (MS). Light particles also include neutrons produced by accelerators. In the following, the discovery of each rubidium isotope is discussed in detail.

\subsection{$^{73}$Rb}\vspace{0.0cm}

In the 1993 publication ``Beta-delayed proton decay of $^{73}$Sr'' Batchelder et al. reported the observation of a proton-unbound excited state of $^{73}$Rb \cite{1993Bat02}. A natural calcium target was bombarded with a 245~MeV $^{36}$Ar beam from the Berkeley 88-inch cyclotron and $^{73}$Rb was populated by $\beta$-decay of the fusion-evaporation residue $^{73}$Sr. The reaction products were transported with a He-jet system onto a rotating wheel which was viewed by two Si-Si detector telescopes. Beta-delayed protons were recorded: ``A single proton group at a laboratory energy of 3.75$\pm$0.04 MeV has been observed, corresponding to the decay of the T = 3/2 isobaric analog state in $^{73}$Rb to the ground state of $^{72}$Kr.'' Previous unsuccessful searches for proton-bound states showed that $^{73}$Rb is the first rubidium isotope beyond the proton dripline \cite{1977DAu01,1991Moh01,1992Yen01}.

\subsection{$^{74}$Rb}\vspace{0.0cm}

The 1976 paper ``The N = Z nuclide $^{74}$Rb with T, I$^{\pi}$ = 1, 0$^{+}$'' by D'Auria et al. described the discovery of $^{74}$Rb \cite{1977DAu01}. A 600 MeV proton beam induced spallation reactions at CERN. $^{74}$Rb was separated with the ISOLDE-2 on-line facility and activities were measured with a $\beta$-particle telescope. ``Improved experimental techniques have permitted the observation of the new nuclide $^{74}$Rb with a half-life of 64.9$\pm$0.5~ms.'' This half-life measurement is included in the weighted average which constitutes the accepted value of 64.776(30)~ms.

\subsection{$^{75}$Rb}\vspace{0.0cm}

Ravn et al. discovered $^{75}$Rb in ``Short-lived isotopes of alkali and alkaline-earth elements studied by on-line isotope separator techniques'' in 1975 \cite{1975Rav01}. The CERN synchrocyclotron accelerated protons to 600 MeV which produced $^{75}$Rb in spallation reactions. The spallation products were separated at the ISOLDE facility. ``The following half-lives of new nucleides have been determined: $^{75}$Rb (21$\pm$3)~sec,...'' This half-life agrees with the presently adopted value of 19.0(12)~s.

\subsection{$^{76}$Rb}\vspace{0.0cm}

The discovery of $^{76}$Rb was reported in ``Identification of new neutron-deficient nuclides $^{76}$Rb and $^{118}$Cs. Half-lives of $^{78}$Rb, $^{119-124}$Cs, $^{126}$Cs'' by Chaumont et al. in 1969 \cite{1969Cha01}. 24 GeV protons from the CERN proton synchrotron bombarded tantalum targets. Rubidium and cesium ions were selectively emitted by surface ionization and separated with an on-line mass spectrometer. ``In a corresponding mass scan in the Rb region $^{76}$Rb was detected as a product of the spallation of Ta. The sensitivity of the apparatus in its present state was not adequate to extend this search to $^{75}$Rb and $^{117}$Cs. The measured cross section for forming $^{76}$Rb is 5$\pm$2 $\mu$b.''

\subsection{$^{77}$Rb}\vspace{0.0cm}

In 1972, Arlt et al. identified $^{77}$Rb in ``Isobars from the spallation reaction for nuclear spectroscopy'' \cite{1972Arl02}. A zirconium-niobium alloy target was bombarded with 660 MeV protons from the Dubna synchrocyclotron. $^{77}$Rb was separated with an electromagnetic isotope separator attached to a surface ionization ion source and identified by measuring $\beta$- and $\gamma$-ray activities with the YASNAPP facility. ``When the zirconium-niobium target was treated as described above a new $\beta$-activity on the mass position A=77 with half-life T$_{1/2}$ = 3.9$\pm$0.1 min has been discovered.'' This half-life agrees with the presently accepted value of 3.77(4)~min. About a year earlier Velandia et al. reported a half-life of 6.1(5)~min \cite{1972Vel01}, however, these results were discredited by de Boer et al. only a month after the paper by Arlt et al.: ``The most embarrassing point is that a 146.5 keV gamma ray which according to Borchert \cite{1971Bor02} should be 31 times stronger than the 106.2 keV gamma ray, is not seen by Velandia et al.'' \cite{1972deB01}. Also, a previously reported half-life of 2.8(10)~min was assigned to either $^{77}$Rb or a $^{77}$Kr isomer \cite{1971Dor01}.

\subsection{$^{78}$Rb}\vspace{0.0cm}

Toeset et al. identified $^{78}$Rb in the 1968 paper ``The decay of rubidium-78 and rubidium-79'' \cite{1968Toe01}. The synchrocyclotron at the Amsterdam Instituut voor Kernphysisch Onderzoek was used to bombard NH$_{4}$Br targets with 36$-$72~MeV $^{3}$He. Gamma-ray spectra were measured with a Ge(Li) detector following chemical separation. ``However, the main argument for concluding that we are dealing with $^{78}$Rb is found in the $\gamma$-spectrum of the 6 minutes period. It is evident from these data that the lowest excited state in the nuclide produced by this decay process is at 455.2 keV, which is in excellent agreement with the observation at 0.45 MeV of the energy of the 2$^{+}$ level produced in $^{78}$Kr by ($\alpha$,$\alpha$') coulomb excitation.'' This measurement corresponds to the 5.74(5)~min isomeric state.

\subsection{$^{79}$Rb}\vspace{0.0cm}

In the 1957 paper ``Mise en evidence d'un isotope leger de rubidium'' Chaminade et al. reported the existence of $^{79}$Rb \cite{1957Cha02}. Fusion evaporation reactions with beams of $^{14}$N, $^{16}$O, and $^{20}$Ne from the cyclotrons of Saclay and Stockholm were used to produce $^{79}$Rb. ``It is shown that the reactions Cu+O, Cu+Ne,Ga+N lead to the formation of a new isotope of Rb, a $\beta^+$ emitter of period 24$\pm$1 min. emitting a complex $\gamma$-radiation of about 150 keV; its mass number is probably 79.'' This half-life is consistent with the presently adopted value of 22.9(5)~min. It is interesting to note that very similar information was included in three additional papers submitted on the same date by various coauthors from Saclay and Stockholm and published in the same issue \cite{1957Bey02,1957Ate01,1957Cru01}. We credit Chaminade et al. with the discovery because it is the one in subsequent publications about $^{79}$Rb \cite{1961Cha01,1968Toe01} and the ENSDF \cite{2008ENS01}.

\subsection{$^{80}$Rb}\vspace{0.0cm}

In 1961 Hoff et al. described the discovery of $^{80}$Rb in ``The neutron deficient chain $^{80}$Sr$-^{80}$Rb'' \cite{1961Hof01}. Ga$_2$O$_3$ targets were irradiated with a 133~MeV $^{14}$N beam from the Berkeley HILAC. $^{80}$Rb was identified with a time-of-flight isotope separator following chemical separation. ``Strontium-80 decays with a half-life 100$\pm$6 min to 34$\pm$4 second $^{80}$Rb, a genetic relationship verified by rapid chemical separation procedures.'' This half-life is included in the calculation of the accepted value of 33.4(70)~s.

\subsection{$^{81,82}$Rb}\vspace{0.0cm}

The observation of $^{81}$Rb and $^{82}$Rb was reported in the 1949 publication ``Mass spectrographic assignment of rubidium isotopes'' by Reynolds et al. \cite{1949Rey01}. The Berkeley 60 and 184-inch cyclotrons accelerated helium ions to energies of 20-100 MeV which irradiated ammonium bromide targets. $^{81}$Rb and $^{82}$Rb were separated with a mass spectrograph and activities were measured with a geiger counter. ``With 80-MeV helium ions, 5.0-hour Rb$^{81}$ predominated in the mixture, and with 20-MeV helium ions almost pure 6.3-hour Rb$^{82}$ was obtained. Otherwise, the similar half-lives would have made characterization of the radiations, which are listed in [the table], very difficult.'' These half-lives agree with the currently accepted values of 4.572(4)~h and 6.472(6)~h for the ground state of $^{81}$Rb and the isomeric state of $^{81}$Rb, respectively.

\subsection{$^{83}$Rb}\vspace{0.0cm}

``Radioactive isotopes of rubidium'' was published by Karraker and Templeton in 1950 describing the discovery of $^{83}$Rb \cite{1950Kar01}. Helium ions accelerated to 18-100 MeV by the Berkeley 60-inch and 184-inch cyclotrons irradiated bromine targets. $^{83}$Rb was separated with a mass spectrograph and activities were measured with a $\beta$-ray spectrometer and end-window Geiger-M\"uller counters following chemical separation. ``The half-life of Rb$^{83}$ was determined by following the decay of a sample of rubidium activity produced by 60-Mev helium ion bombardment of bromine, very similar to the sample which gave a radioactive transfer at masses 83 and 84. Resolution of the decay curve showed a 107-day activity, assigned to mass 83, as well as the 34-day Rb$^{84}$.'' This half-life is reasonably close to the accepted value of 86.2(1)~d.

\subsection{$^{84}$Rb}\vspace{0.0cm}

Barber reported first evidence for $^{84}$Rb in the 1947 article ``Search for positron-electron branching in certain beta emitting isotopes'' \cite{1947Bar01}. Rubidium and strontium was bombarded with neutrons and 18 MeV deuterons, respectively produced by the 60-inch Berkeley cyclotron. The activated rubidium samples were measured with a trochoid with a Geiger M\"uller counter and the activities from the strontium samples were measured following chemical separation. ``The rubidium was chemically separated from the other reaction products, and the half-life of the positron activity was measured roughly as 40 days. These two methods of production suggest that the activity is due to Rb$^{84}$, which could decay to Kr$^{84}$ by positron emission, from the reactions Rb$^{85}$(n,2n)$\rightarrow$Rb$^{84}$, Sr$^{86}$(d,$\alpha$)$\rightarrow$Rb$^{84}$.'' This half-life is consistent with the currently accepted value of 32.82(14)~d.

\subsection{$^{85}$Rb}\vspace{0.0cm}

The discovery of stable $^{85}$Rb was reported by Aston in his 1921 paper ``The constitution of the alkali metals'' \cite{1931Ast03}. The positive anode ray method was used to identify $^{85}$Rb with the Cavendish mass spectrograph. ``Rubidium (atomic weight 85.45) gives two lines two units apart of relative intensity about 3 to 1. Comparison with the potassium line 39 gives these the masses 85 and 87 to within a fraction of a unit.''

\subsection{$^{86}$Rb}\vspace{0.0cm}

The 1941 paper ``Radioactive Rb from deuteron bombardment of Sr'' by Helmholz et al. documented the identification of $^{86}$Rb \cite{1941Hel01}. Deuterons with an energy of 16 MeV from the Berkeley 60-inch cyclotron irradiated strontium targets. The resulting activities were measured following chemical separation. ``Of the two slow neutron activities, only that of Rb$^{86}$ can be produced by the Sr(d,$\alpha$)Rb reaction, since Sr$^{88}$ is the heaviest stable isotope of Sr. Therefore this 19.5-day period can definitely be assigned to Rb$^{86}$.'' This value agrees with the currently accepted value of 18.642(18)~d. An 18~d activity had been reported earlier, but no definite mass assignment was made \cite{1937Sne01}.

\subsection{$^{87}$Sr}\vspace{0.0cm}

The discovery of stable $^{87}$Rb was reported by Aston in his 1921 paper ``The constitution of the alkali metals'' \cite{1931Ast03}. The positive anode ray method was used to identify $^{87}$Rb with the Cavendish mass spectrograph. ``Rubidium (atomic weight 85.45) gives two lines two units apart of relative intensity about 3 to 1. Comparison with the potassium line 39 gives these the masses 85 and 87 to within a fraction of a unit.''

\subsection{$^{88}$Rb}\vspace{0.0cm}

The identification of $^{88}$Rb was reported in 1939 by Heyn et al. in the article ``Transmutation of uranium and thorium by neutrons'' \cite{1939Hey01}. An uranyl nitrate solution was irradiated with a strong neutron source of the Philips X-Ray Laboratory. Resulting activities were measured following chemical separation. ``Active krypton forms radio-rubidium with a period of 16$\pm$2 minutes. Probably this period is due to the known radioactive rubidium isotope obtained from rubidium by neutron capture. The half-life of the latter we found to be 17$\pm$2 minutes. In this connexion, it may be mentioned that from uranium bombarded with neutron Curie and Savitch obtained precipitates of alkali salts showing a period of 18 minutes. Based on the experiments described we suggest the following processes: ...  $^{88}$Kr$\rightarrow^{88}$Rb(17m.)$\rightarrow^{88}$Sr(stable),'' This half-life agrees with the currently accepted value of 17.773(11)~min. The 18~min half-life mentioned in the quote was published in reference \cite{1939Cur01}.

\subsection{$^{89}$Rb}\vspace{0.0cm}

Glasoe and Steigman identified $^{89}$Rb in the 1940 paper ``Radioactive products from gases produced in uranium fission'' \cite{1940Gla02}. Uranium nitrate was irradiated with neutrons produced by bombarding beryllium with protons from the cyclotron at Columbia University. The resulting activities were measured with Geiger M\"uller counters following chemical separation. ``A typical decay curve for the material obtained in this manner is shown in [the figure] with the indicated half-life of 15.4$\pm$0.2 minutes. The active Sr into which this Rb decays was obtained under similar irradiation and collection conditions and was found to have a period of 51$\pm$2 days. It may be assumed that this is the same Sr isotope reported by Stewart and by DuBridge and Marshall as obtained from Sr(d,p) and assigned to Sr$^{89}$.'' The measured half-life of 15.4(2)~min for $^{89}$Rb agrees with the presently adopted value of 15.15(12)~min. Stewart \cite{1937Ste01,1939Ste01} and DuBridge and Marshall \cite{1939DuB01} had measured 55~d for the half-life of $^{89}$Sr earlier, thus allowing the assignment of the activity observed by Glasoe and Steigman to $^{89}$Rb.

\subsection{$^{90,91}$Rb}\vspace{0.0cm}

Kofoed-Hansen and Nielsen reported the discovery of  $^{90}$Rb and $^{91}$Rb in the 1951 paper ``Short-lived krypton isotopes and their daughter substances'' \cite{1951Kof01}. Uranium was bombarded with neutrons produced at the Copenhagen cyclotron and fission fragments were transported to an ion source of a mass separator. Activities were measured following chemical separation. ``Furthermore it was found that Rb$^{91}$ has an isomeric state. Both these states of Rb$^{91}$ (half-lives 100 sec and 14 min, respectively) decay to the well-known Sr$^{91}$ 9.7 hr which again decays to the 60-day and the 50-min isomers of Y$^{91}$.'' The half-life of 2.74~min for $^{90}$Rb was listed in a table. The half-life of 2.74~min for $^{90}$Rb agrees with the currently accepted values of 158(5)~s. Although the half-life of 100~s for $^{91}$Rb differs from the accepted value of 58.4(4)~s by almost a factor of two and the existence of a 14~min isomer was found to be incorrect by Wahl et al. \cite{1962Wah01} we credit the discovery to Kofoed-Hansen and Nielsen, because Wahl et al. did not question the validity of the ground-state reporting a 72~s half-life. An 80~s half-life was observed earlier without a mass assignment \cite{1940Hah02}.

\subsection{$^{92,93}$Rb}\vspace{0.0cm}

The discovery of $^{92}$Rb and $^{93}$Rb was described in the 1960 article ``The identification and half lives of fission-Product $^{92}$Rb and $^{93}$Rb'' by Fritze and Kennett \cite{1960Fri01}. A uranyl nitrate solution, enriched to 93\%  $^{235}$U, was irradiated at the McMaster University research reactor and decay curves were measured with a $\beta$-scintillation detector following chemical separation. ``The existence of two new rubidium isotopes, Rb$^{92}$ and Rb$^{93}$, has been established and their half lives measured.'' The half lives of these short-lived fission products were determined using a technique of timed precipitations. The values obtained for Rb$^{92}$ and Rb$^{93}$ were 5.3$\pm$0.5 sec and 5.6$\pm$0.5 sec, respectively. These half-lives agree with the presently accepted values of 4.492(20)~s and 5.84(2)~s for $^{92}$Rb and $^{93}$Rb, respectively.

\subsection{$^{94}$Rb}\vspace{0.0cm}

$^{95}$Sr was observed for the first time as reported in the 1961 paper ``Half-lives of Rb$^{94}$, Sr$^{94}$, Y$^{94}$, Rb$^{95}$, Sr$^{95}$, Y$^{95}$'' by Fritze et al. \cite{1961Fri01}. A $^{235}$U solution was irradiated in the McMaster University research reactor and $^{94}$Rb was identified by timed precipitations. ``The existence of a new rubidium isotope, Rb$^{94}$, has been established and its half-life measured. The half-life of this fission product was determined using the technique of timed precipitations. The value obtained for Rb$^{94}$ was 2.9$\pm$0.3 seconds.'' This value agrees with the presently adopted value of 2.702(5)~s.

\subsection{$^{95,96}$Rb}\vspace{0.0cm}

Amarel et al. observed $^{95}$Rb and $^{96}$Rb in 1967 as reported in their article ``Half life determination of some short-lived isotopes of Rb, Sr, Cs, Ba, La and identification of $^{93,94,95,96}$Rb as delayed neutron precursors by on-line mass-spectrometry'' \cite{1967Ama01}. The isotopes were produced by fission of $^{238}$U induced by 150 MeV protons from the Orsay synchrocyclotron. $^{95}$Rb and $^{96}$Rb were identified with a Nier-type mass spectrometer and half-lives were determined by $\beta$ counting with a plastic scintillator telescope. The measured half-lives were listed in a table with 360(20)~ms for $^{95}$Rb and 230(20)~ms for $^{96}$Rb. These half-lives are consistent with the currently adopted values of 377.5(8)~ms and 203(3)~ms for $^{95}$Rb and $^{96}$Rb, respectively. A previous study was only able to set an upper limit of 2.5~s on the $^{95}$Rb half-life \cite{1961Fri01}.

\subsection{$^{97}$Rb}\vspace{0.0cm}

In 1969, the article ``Delayed neutron emission probabilities of Rb and Cs precursors. The half-life of $^{97}$Rb'' by Amarel et al. described the first observation of $^{97}$Rb \cite{1969Ama01}. Protons accelerated to 150 MeV by the Orsay synchrocyclotron bombarded $^{238}$U targets. $^{97}$Rb was identified with an on-line mass spectrometer and  $^{10}$BF$_{3}$ neutron counters. ``The rubidium isotope of mass 97 has been identified as a precursor and its period (0.135 sec) determined through observation of the delayed neutrons.'' This half-life is consistent with the currently accepted value of 169.1(6)~ms.

\subsection{$^{98,99}$Rb}\vspace{0.0cm}

$^{98}$Rb and $^{99}$Rb were first observed in 1971 by Tracy et al. in the article ``Half-lives of the new isotopes $^{99}$Rb, $^{98}$Sr, $^{145-146}$Cs'' \cite{1971Tra01}. Fission fragments from the bombardment of 50 MeV protons on $^{238}$U at the Grenoble cyclotron were studied. Beta-particles were measured at the end of an on-line mass spectrometer: ``On-line mass spectrometer techniques for the separation of Rb and Cs have been used to detect products from the 50 MeV proton induced fission of $^{238}$U. The new isotopes $^{99}$Rb, $^{98}$Sr and $^{145-146}$Cs were observed and their half-life measured. Also the half-life of $^{98}$Rb was measured for the first time.'' The measured half-lives of 136(8)~ms and 76(5)~ms were listed in a table and are close to the presently adopted values of 114(5)~ms and 50.3(7)~ms for $^{98}$Rb and $^{99}$Rb, respectively. The observation of $^{98}$Rb was not considered a discovery because of a previous publication in a conference proceeding \cite{1966Ama01}.

\subsection{$^{100}$Rb}\vspace{0.0cm}

Koglin et al. reported the identification of $^{100}$Rb in ``Half-lives of the new isotopes $^{100}$Rb, $^{100}$Sr, $^{148}$Cs and of $^{199}$Rb, $^{99}$Sr and $^{147}$Cs'' in 1978 \cite{1978Kog01}. $^{100}$Rb was produced and identified by neutron induced fission of $^{235}$U at the On-line Separator f\"{u}r Thermisch Ionisierbare Spaltprodukte (OSTIS) facility in Grenoble, France. ``An improvement of the ion source of the online fission product separator OSTIS allowed us to identify the new isotopes $^{100}$Rb (50$\pm$10~msec), $^{100}$Sr (170$\pm$80~ms), and $^{148}$Cs (130$\pm$40~ms).'' This half-life of $^{100}$Rb is included in the weighted average of the accepted value of 51(8)~ms. The observation of $^{100}$Rb was independently reported less than a month later by Peuser et al. \cite{1979Peu01}.

\subsection{$^{101}$Rb}\vspace{0.0cm}

Balog et al. published the discovery of $^{101}$Rb in ``Experimental beta-decay energies of very neutron-rich isobars with mass numbers A=101 and A=102'' in 1992 \cite{1992Bal01}.``The Q$_{\beta}$-values of $^{101}$Rb, $^{101,102}$Sr and $^{101,102}$Y have been measured for the first time at a mass separator ISOLDE by means of $\beta$$\gamma$-coincidence techniques with a plastic scintillation detector telescope and a large Ge(HP)-detector.'' The half-life for $^{101}$Rb was measured to be 32~ms and listed in a table. This value agrees with the presently accepted value of 32(5)~ms.

\subsection{$^{102}$Rb}\vspace{0.0cm}

The first observation of $^{102}$Rb was described by Lhersonneau et al. in the 1995 article ``First evidence for the 2$^+$ level in the very neutron-rich nucleus $^{102}$Sr'' \cite{1995Lhe01}. A UC target was bombarded with 1 GeV protons from the PS booster at CERN. $^{102}$Rb was identified with the General Purpose Separator at ISOLDE and activities were measured with two Ge detectors and a thin plastic scintillator. ``[The figure] suggests that the 126 keV line is neither produced in the $^{102}$Sr nor $^{83}$Sr decays. Since such a line has not been reported in any other activities identified by us, it was assigned to the $\beta$-decay of $^{102}$Rb into the 2$^+$ state of $^{102}$Sr.'' In their measurement Lhersonneau et al. used the half-life value of 37~ms reported only in a conference proceeding \cite{1987Pfe01}. The half-life of $^{102}$Rb has still not been published in the refereed literature.

\subsection{$^{103}$Rb}\vspace{0.0cm}

The discovery of $^{103}$Rb was reported in the 2010 article ``Identification of 45 new neutron-rich isotopes produced by in-flight fission of a $^{238}$U Beam at 345 MeV/nucleon,'' by Ohnishi et al. \cite{2010Ohn01}. The experiment was performed at the RI Beam Factory at RIKEN, where the new isotopes were created by in-flight fission of a 345 MeV/nucleon $^{238}$U beam on a beryllium target. $^{103}$Rb was separated and identified with the BigRIPS superconducting in-flight separator. The results for the new isotopes discovered in this study were summarized in a table. Ninety-nine counts were recorded for $^{103}$Rb.

\section{Discovery of $^{73-107}$Sr}

Thirty-five strontium isotopes from A = 73--107 have been discovered so far; these include 4 stable, 12 neutron-deficient and 19 neutron-rich isotopes. According to the HFB-14 model \cite{2007Gor01}, $^{119}$Sr should be the last odd-even particle stable neutron-rich nucleus while the even-even particle stable neutron-rich nuclei should continue through $^{130}$Sr. At the proton dripline three more isotopes $^{70-72}$Sr should be particle stable and in addition $^{69}$Sr and $^{68}$Sr could have half-lives longer than 10$^{-21}$~s \cite{2004Tho01}. Thus, about 23 isotopes have yet to be discovered corresponding to 40\% of all possible strontium isotopes.

\begin{figure}
	\centering
	\includegraphics[scale=.5]{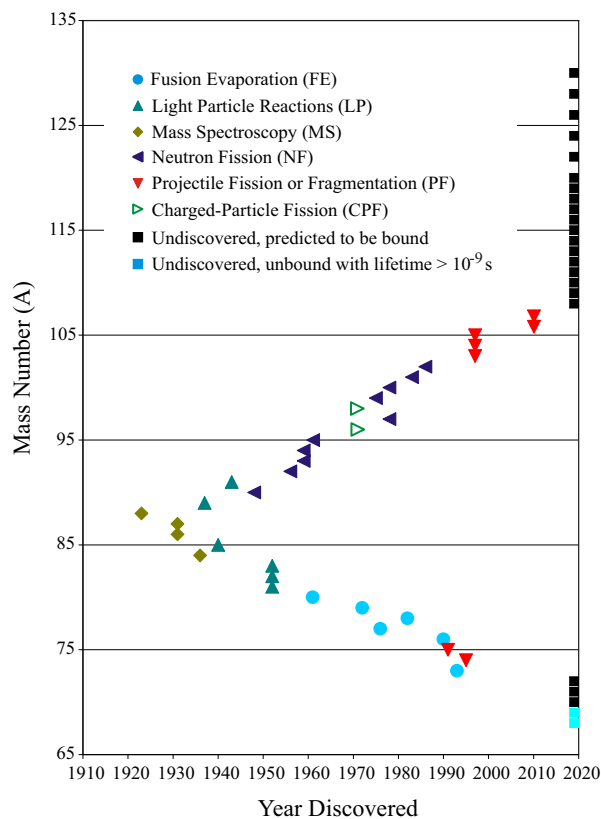}
	\caption{Strontium isotopes as a function of time when they were discovered. The different production methods are indicated. The solid black squares on the right hand side of the plot are isotopes predicted to be bound by the HFB-14 model. On the proton-rich side the light blue squares correspond to unbound isotopes predicted to have lifetimes larger than $\sim 10^{-9}$~s.}
\label{f:year-strontium}
\end{figure}

Figure \ref{f:year-strontium} summarizes the year of first discovery for all strontium isotopes identified by the method of discovery. The range of isotopes predicted to exist is indicated on the right side of the figure. The radioactive strontium isotopes were produced using fusion-evaporation reactions (FE), light-particle reactions (LP), neutron (NF) and charged-particle induced (CPF) fission, and projectile fragmentation or projectile fission (PF). The stable isotopes were identified using mass spectroscopy (MS). Light particles also include neutrons produced by accelerators. The discovery of each strontium isotope is discussed in detail and a summary is presented in Table 1.

\subsection{$^{73}$Sr}\vspace{0.0cm}

In the 1993 publication ``Beta-delayed proton decay of $^{73}$Sr'' Batchelder et al. reported the observation of $^{73}$Sr \cite{1993Bat02}. A natural calcium target was bombarded with a 245~MeV $^{36}$Ar beam from the Berkeley 88-inch cyclotron and $^{73}$Sr was produced in the fusion-evaporation reaction $^{40}$Ca($^{36}$Ar,3n). The reaction products were transported with a He-jet system onto a rotating wheel which was viewed by two Si-Si detector telescopes. Beta-delayed protons were recorded: ``This spectrum clearly shows a peak containing 21 counts at 3.77$\pm$0.05 MeV, which, as discussed below, we assign to the $\beta$-delayed proton decay of $^{73}$Sr.'' 

\subsection{$^{74}$Sr}\vspace{0.0cm}

In the 1995 article ``New isotopes from $^{78}$Kr fragmentation and the ending point of the astrophysical rapid-proton-capture process'' Blank et al. reported the discovery of $^{74}$Sr \cite{1995Bla01}. A 73~MeV/nucleon $^{78}$Kr beam bombarded a nickel target \cite{1995Bla01} at GANIL. $^{74}$Sr was produced via projectile fragmentation and identified with the SISSI/LISE facility by measuring the time-of-flight through the separator and the $\Delta$E-E in a silicon detector telescope. A lower limit for the half-life was established, ``We find clear evidence for $^{60}$Ga, $^{64}$As, $^{69,70}$Kr, and $^{74}$Sr.''

\subsection{$^{75}$Sr}\vspace{0.0cm}

Mohar et al. first observed $^{75}$Sr in the 1991 paper ``Identification of new nuclei near the proton-dripline for 31$\leq$Z$\leq$38'' \cite{1991Moh01}. A 65 A$\cdot$MeV $^{78}$Kr beam produced by the Michigan State K1200 cyclotron reacted with an enriched $^{58}$Ni target. $^{75}$Sr was identified by measuring the rigidity, $\Delta$E, E$_{total}$, and velocity in the A1200 fragment separator. ``Several new isotopes at or near the proton-drip line are indicated in the mass spectra: $^{61}$Ga, $^{62}$Ge, $^{63}$Ge, $^{65}$As, $^{69}$Br, and $^{75}$Sr.''

\subsection{$^{76}$Sr}\vspace{0.0cm}

In 1990, Lister et al. reported the observation of $^{76}$Sr in the paper ``Shape changes in N=Z nuclei from germanium to zirconium'' \cite{1990Lis01}.  A 175 MeV $^{54}$Fe beam from the Daresbury NSF tandem accelerator bombarded an enriched $^{24}$Mg target and $^{76}$Sr was produced in the fusion-evaporation reaction $^{24}$Mg($^{54}$Fe,2n). Gamma-rays were measured with ten shielded germanium detectors in coincidence with recoil products recorded in an ionization chamber. ``In the $^{76}$Sr experiment, the identification of the 260.9$\pm$0.2 keV first excited state was particularly straightforward as no other $\gamma$ rays near this energy were produced in the fusion of $^{54}$Fe and $^{24}$Mg.''

\subsection{$^{77}$Sr}\vspace{0.0cm}

Hardy et al. published the discovery of $^{77}$Sr in the 1976 paper ``A new series of beta-delayed proton precursors'' \cite{1976Har01}. A 130~MeV $^{40}$Ca beam from the Chalk River MP tandem accelerator was incident upon a calcium target and $^{77}$Sr was produced in the fusion-evaporation reaction $^{40}$Ca($^{40}$Ca,2pn). Beta-delayed protons were recorded with a surface barrier counter telescope; in addition X-rays and $\gamma$-rays were measured in coincidence with a NaI(Tl) detector. ``The half-life of the previously unreported $^{77}$Sr is 9.0$\pm$1.0 seconds.'' This half-life agrees with the currently accepted value of 9.0(2)~s.

\subsection{$^{78}$Sr}\vspace{0.0cm}

``Extreme prolate deformation in light strontium isotopes'' by Lister et al. identified $^{78}$Sr in 1982 \cite{1982Lis01}. A $^{58}$Ni target was bombarded with a 100~MeV beam of $^{24}$Mg from the Brookhaven Van de Graaff accelerator facility, and $^{78}$Sr was formed in the fusion-evaporation reaction $^{58}$Ni($^{32}$Mg,2p2n). The isotope was identified by triple coincidence measurements of light-charged particles, neutrons and $\gamma$-rays. ``Levels in $^{78}$Sr were seen to J=10 with E(2$^{+}$) = 278 keV and T$_{1/2}$ = 155$\pm$19 ps.'' The first half-life measurement of the $^{78}$Sr ground state was reported only five months later \cite{1982Lia01}. A previously reported half-life of 30.6(23)~min \cite{1971Bil01} was evidently incorrect.

\subsection{$^{79}$Sr}\vspace{0.0cm}

Landenbauer-Bellis et al. reported the observation of $^{79}$Sr in the 1972 paper ``Energies and intensities of $\gamma$ rays in the decay of $^{79}$Sr'' \cite{1972Lad01}. A 65~MeV $^{14}$N beam from the Yale Heavy Ion Accelerator activated enriched $^{69}$Ga targets and $^{79}$Sr was formed in the fusion-evaporation reaction $^{69}$Ga($^{14}$N,4n). Gamma-ray spectra of the activated samples were recorded with high-resolution Ge(Li) detectors. ``Our conclusions are that $^{79}$Sr has a half-life of 4.4$\pm$0.2 minutes associated with a single $\gamma$-ray having the energy of 105.4$\pm$0.2 keV.'' This half-life is almost a factor two larger than the presently accepted value of 2.25(10)~min, however, the $\gamma$-ray decay of the second excited state in the daughter nucleus $^{79}$Rb at 105.4(2)~keV was identified correctly. A previously reported half-life of 8.1(3)~min. \cite{1971Bil02} was evidently incorrect. In 1971, Doron and Blann observed a half-life of 1.9~min in coincidence with three $\gamma$-rays consistent with the decay of $^{79}$Sr, however, they argued that it could not be $^{79}$Sr because of a growth in the activity of $^{79}$Kr \cite{1971Dor01}.

\subsection{$^{80}$Sr}\vspace{0.0cm}

In 1961 Hoff et al. described the discovery of $^{80}$Sr in ``The neutron deficient chain $^{80}$Sr$-^{80}$Rb'' \cite{1961Hof01}. Ga$_2$O$_3$ targets were irradiated with a 133~MeV $^{14}$N beam from the Berkeley HILAC. $^{80}$Sr was identified with a time-of-flight isotope separator following chemical separation. ``Strontium-80 decays with a half-life 100$\pm$6 min to 34$\pm$4 second $^{80}$Rb, a genetic relationship verified by rapid chemical separation procedures.'' This half-life for $^{80}$Sr agrees with the currently adopted value of 106.3(15)~min.

\subsection{$^{81-83}$Sr}\vspace{0.0cm}

$^{81}$Sr, $^{82}$Sr and $^{83}$Sr were observed by Castner and Templeton in ``Some neutron deficient strontium isotopes'' published in 1952 \cite{1952Cas01}. The Berkeley 184-in. synchrocyclotron accelerated protons to 25$-$100 MeV which then irradiated a rubidium chloride target. Activities were recorded as a function of time following chemical separation. ``Bombardment of rubidium with protons produced $^{81}$Sr, half-life 29 minutes, $^{82}$Sr, half-life 25 days, $^{83}$Sr, half-life 38 hours.'' These half-lives are consistent with the presently accepted values of 22.3(4)~min, 25.36(3)~d, 32.41(3)~h, for $^{81}$Sr, $^{82}$Sr, and $^{83}$Sr, respectively. Caretto and Wiig submitted their results on $^{82}$Sr and $^{83}$Sr only a week later \cite{1952Car01}.

\subsection{$^{84}$Sr}\vspace{0.0cm}

In 1936 Blewett and Sampson reported the discovery of stable $^{84}$Sr in ``Isotopic constitution of strontium, barium, and indium'' \cite{1936Ble01}. The mass spectrographic study of strontium was performed at Princeton University by heating strontium oxide from a tungsten filament. ``In the case of strontium, a peak was observed at mass 84 whose height corresponded to 0.5 percent of the total strontium emission. The masses of the known isotopes of strontium are all greater than 84 so the possibility is excluded that the effect is due to a strontium compound. Since the only known isotope of mass 84 belongs to krypton and the other krypton isotopes do not appear, we believe that this peak is due to a new isotope of strontium.''

\subsection{$^{85}$Sr}\vspace{0.0cm}

The first observation of $^{85}$Sr was documented by DuBridge and Marshall in their paper 1940 paper entitled ``Radioactive isotopes of Sr, Y, and Zr'' \cite{1940DuB01}. Rubidium targets were bombarded with 6.7~MeV protons at the University of Rochester. The activities of the reaction products were measured with a freon-filled ionization chamber, and the $\beta$ and $\gamma$ decays were measured with a magnetic cloud chamber and a $\beta$-ray spectrograph. ``There are only two stable isotopes of Rb, of masses 85 and 87, the latter being naturally $\beta$-active with a period of 10$^{10}$ years. Rb(p,n) reactions therefore would yield only Sr$^{85}$ and Sr$^{87}$. Since Sr$^{87}$ is stable it would appear necessary at first sight to assign all three periods to Sr$^{85}$. However it will be shown below that the 2.75-hr. period must be assigned to a metastable state of Sr$^{87}$. The 70-min. and 66-day periods are therefore assigned to Sr$^{85}$.'' The 66-day half-life agrees with the currently accepted value of 64.853(8)~d. The 70-min half-life corresponds to an isomeric state.

\subsection{$^{86}$Sr}\vspace{0.0cm}

In 1924 Aston reported the first observation of stable $^{86}$Sr in ``Recent results obtained with the mass-spectrograph'' \cite{1924Ast04}. The mass spectra were measured at the Cavendish Laboratory in Cambridge, UK: ``Experiments with strontium have been repeated under better conditions. The conclusion that it consists almost entirely of atoms of mass number 88 has been confirmed, but the greater sharpness of the mass spectra has now revealed two points which explain to some extent the low value 87.62 assigned to its chemical atomic weight. In the first place, a very faint constituent at 86 has been discovered which appears to be present to the extent of 3 or 4 percent.''

\subsection{$^{87}$Sr}\vspace{0.0cm}

Stable $^{87}$Sr was reported by Aston in his 1931 letter ``New isotopes of strontium and barium'' \cite{1931Ast01}. The mass spectra were measured at the Cavendish Laboratory in Cambridge, UK: ``Application of high resolution to accelerated anode rays has now provided improved mass spectra of these two elements. Strontium shows a third isotope 87 in addition to those already observed.''

\subsection{$^{88}$Sr}\vspace{0.0cm}

The first stable strontium isotope was discovered in 1923 by Aston in ``Further determinations of the constitution of the elements by the method of accelerated anode rays'' \cite{1923Ast01}. The isotope was identified by its mass spectra recorded at the Cavendish Laboratory in Cambridge, UK. ``The mass spectrum of strontium shows one line only, at 88. This was obtained in considerable intensity. If any other constituents exist they must be present in very small quantities, so that it is practically certain that the chemical atomic weight 87.63 at present in use is too low.''

\subsection{$^{89}$Sr}\vspace{0.0cm}

Stewart et al. discovered $^{89}$Sr, which was reported in the 1937 publication ``Induced radioactivity in strontium and yttrium'' \cite{1937Ste01}. Deuterons of 6.3~MeV from the University of Michigan cyclotron bombarded strontium targets to produce the new isotope. Decay curves were recorded with a Lauritsen type electroscope or a Wulf string electrometer following chemical separation. ``From deuteron bombardment of strontium, two definitive negative-active periods were obtained. These were found in every strontium precipitate, and were found to have half-lives of 3.0$\pm$0.1 hours and 55$\pm$5 days.'' The second half-life agrees with the currently accepted value of 50.53(7)~d, while the shorter half-life most likely corresponds to an isomeric state of stable $^{87}$Sr.

\subsection{$^{90}$Sr}\vspace{0.0cm}

In the 1948 paper ``Fission products of U$^{233}$,'' Grummitt and Wilkinson identified $^{90}$Sr \cite{1948Gru01}. Fission products from neutron irradiation of $^{233}$U were measured at Chalk River, Canada. Gamma-ray  and $\beta$-ray absorption data as well as half-life measurements were performed following chemical separation. No experimental details are given and the results are summarized in a table. The reported half-life of $\sim$30~y is consistent with the presently accepted value of 28.79(6)~y. In an earlier paper Grummitt and Wilkinson had reported a half-life of  $\sim$70~y \cite{1946Gru01}. Six month later Hayden identified $^{90}$Sr independently with a mass spectrometer \cite{1948Hay01}. In 1943, Hahn and Strassmann quoted only a lower limit of 5 years for the half-life of $^{90}$Sr \cite{1943Hah01}.

\subsection{$^{91}$Sr}\vspace{0.0cm}

$^{91}$Sr was first correctly identified by Seelmann-Eggebert in ``\"Uber einige aktive Yttrium-Isotope'' in 1943 \cite{1943See02}. Zirconium was irradiated with fast Li/D neutrons and several yttrium activities were measured following chemical separation. ``Es konnte festgestellt werden, da\ss\ das mit dem 57 Tage Yttrium isomere 50 Minuten-Yttrium der angeregte Zustand ist, und da\ss\ die Muttersubstanz dieser Isomeren, das 10 Stunden-Strontium, auch bei Bestrahlung des Zirkons mit schnellen Neutronen durch n,$\alpha$ Proze\ss\ entsteht. Die Masse dieser Reihe ist daher 91.'' [It could be determined that the 50-min yttrium isomer of the 57-day yttrium corresponds to the excited state and that the 10-hour strontium mother substance of these isomers could also be produced in the n,$\alpha$ process by bombarding zirconium with fast neutrons. The mass of this chain is thus 91.] The 10-hr half-life measured for $^{91}$Sr agrees with the presently accepted value of 9.63(5)~h. G\"otte had previously linked a 8.5~h half-life to the 50-min. yttrium, however, without a mass assignment \cite{1941Got01}.

\subsection{$^{92}$Sr}\vspace{0.0cm}

Herrmann and Strassmann observed $^{92}$Sr in the 1956 paper ``\"Uber einige Strontium- und Yttrium Isotope bei der Uranspaltung'' \cite{1956Her01}. Fission fragments of uranium were produced by neutron irradiation at the Harwell reactor in the UK and by bombardment of slowed-down neutrons from the cascade accelerator at Mainz, Germany. Beta-ray absorption measurements were performed following chemical separation. ``Die Zuordnung des 2.7~h Sr zur Massenzahl 92 beruht auf der Zuordnung des Folgeprodukts $^{92}$Y.'' [The 2.7~h strontium was assigned to the mass number 92 based on the identification of the daughter $^{92}$Y.] The measured half-life of 2.60(4)~h agrees with the currently adopted value of 2.66(4)~h. Strontium activities of 3~h \cite{1937Poo01} and 2.7~h \cite{1941Got01,1943Hah02} had previously been reported without definite mass assignments.

\subsection{$^{93,94}$Sr}\vspace{0.0cm}

The first observation of $^{93}$Sr and $^{94}$Sr was reported in ``Radiations of $^{93}$Y and $^{94}$Y and half-lives of $^{93}$Sr and $^{94}$Sr,'' in 1959 by Knight et al. \cite{1959Kni01}. Neutrons from the Los Alamos Water Boiler reactor irradiated $^{235}$U and the fission fragments were chemically separated and $\beta$-decay curves were recorded. ``The $^{93}$Sr half-life was obtained by measurement of the amount of the 10.25~hr $^{93}$Y component in yttrium samples milked from fission-product strontium as described in the previous section... The $^{93}$Sr half-life obtained from these measurements was 8.2$\pm$0.8~min. The $^{94}$Sr half-life was obtained in a similar manner... the indicated $^{94}$Sr half-life is 1.3$\pm$0.2~min.'' These half-lives are in reasonable agreement with the currently accepted values of 7.423(24)~min and 75.3(20)~s for $^{93}$Sr and $^{94}$Sr, respectively. A 7-min half-life had been reported previously without a mass assignment \cite{1939Lie01} and a 2~min half-life could only be assigned to a mass $>$91 \cite{1943Hah01}.

\subsection{$^{95}$Sr}\vspace{0.0cm}

The discovery of $^{95}$Sr was reported in the 1961 paper ``Half-lives of Rb$^{94}$, Sr$^{94}$, Y$^{94}$, Rb$^{95}$, Sr$^{95}$, Y$^{95}$'' by Fritze et al. \cite{1961Fri01}. A $^{235}$U solution was irradiated in the McMaster University research reactor and $^{95}$Sr was identified by timed precipitations. ``Two values for the half-life of Sr$^{95}$ were obtained from the relative quantities of Y$^{95}$ in the yttrium samples of runs No. 1 and No. 2 which are mentioned in Section (B). The values obtained were: (1) 0.6$_8$ minute, (2) 0.8$_8$ minute, average 0.8$\pm$0.15 minute.'' Although this average half-life differs from the accepted half-life of 23.90(14)~s by about a factor of two, we credit Fritze et al. with the first identification, because subsequent papers first measuring 33(6)~s \cite{1966Nor02} and then 26(1)~s \cite{1967Ama01} did not discredit these earlier results.

\subsection{$^{96}$Sr}\vspace{0.0cm}

In the 1971 article ``Quelques r\'esultats nouveaux sur les noyaux de masse paire de strontium riches en neutrons,'' Macias-Marques et al. described the observation of $^{96}$Sr \cite{1971Mac01}. $^{96}$Sr was observed in the decay of $^{96}$Rb which was produced by proton induced fission of $^{238}$U at the synchrocyclotron of Orsay, France. ``Un premier r\'esultat de ces mesures est la connaissance des \'energies du premier niveau 2$^+$ des Sr$^{92,94,96}$: 813, 835 et 813 keV respectivement.'' [First results of these measurements are the energies of the first excited 2$^+$ of Sr$^{92,94,96}$ which are 813, 835 and 813 keV, respectively.] A previous tentative assignment of 2$^+$ and 4$^+$ energies of 204.1~keV and 556.3~keV \cite{1970Che01} as well as a measurement of the half-life of 4.0(2)~s \cite{1967Ama01} could not be confirmed. Previously measured isomeric excited states at 101.0 and 110.7~keV were assigned to either $^{96}$Sr or $^{97}$Sr \cite{1970Gru01}.

\subsection{$^{97}$Sr}\vspace{0.0cm}

Wohn et al. reported the discovery of $^{97}$Sr in the 1978 article ``Identification of $^{147}$Cs and half-life determinations for Cs and Ba isotopes with A=144-147 and Rb and Sr isotopes with A=96-98'' \cite{1978Woh01}. $^{97}$Sr was produced and identified by neutron induced fission of $^{235}$U at the On-line Separator f\"{u}r Thermisch Ionisierbare Spaltprodukte (OSTIS) facility of the Institut Laue-Langevin in Grenoble, France. ``Half-life determinations of Rb and Cs fission products available at an on-line mass separator have been made for several neutron-rich Rb, Sr, Cs, and Ba isotopes using both $\beta$-multiscale and $\gamma$-multispectra measurements. The half-lives and rms uncertainties (in sec) are ... $^{97}$Sr, 0.441$\pm$0.015.'' This half-life agrees with the presently accepted value of 429$\pm$5~ms. Previously measured isomeric excited states at 101.0 and 110.7~keV were assigned to either $^{96}$Sr or $^{97}$Sr \cite{1970Gru01}.

\subsection{$^{98}$Sr}\vspace{0.0cm}

$^{98}$Sr was first observed in 1971 by Tracy et al. in the article ``Half-lives of the new isotopes $^{99}$Rb, $^{98}$Sr, $^{145-146}$Cs'' \cite{1971Tra01}. Fission fragments from the bombardment of 50 MeV protons on $^{238}$U at the Grenoble cyclotron were studied. Beta-particles were measured at the end of an on-line mass spectrometer. ``At mass 98 only case (i) will account for the observed ratio and we are led to assign the observed 0.845 second activity to $^{98}$Sr.'' This measurement is close to the accepted value of 0.653(2)~s.

\subsection{$^{99}$Sr}\vspace{0.0cm}

The first observation of $^{99}$Sr was described in ``The P$_{n}$ values of the $^{235}$U(n$_{th}$,f) produced precursors in the mass chains 90, 91, 93-95, 99, 134 and 137-139,'' in 1975 by Asghar et al. \cite{1975Asg01}. $^{235}$U targets were irradiated with neutrons from the Grenoble high flux reactor. The Lohengrin mass separator was used to identify the fission fragments by measuring the mass-to-charge ratio as well as the energy distribution and $\beta$-ray activities. ``The present work led to three new periods corresponding to the new isotopes of selenium ($^{91}$Se, T$_{1/2}$ = 0.27$\pm$0.05~sec), strontium ($^{99}$Sr, T$_{1/2}$ = 0.6$\pm$0.2~sec), and tellurium $^{138}$Te, T$_{1/2}$ = 1.3$\pm$0.3~sec).'' This half-life for $^{99}$Sr agrees with the presently accepted value of 0.269(10)~s.

\subsection{$^{100}$Sr}\vspace{0.0cm}

Koglin et al. reported the identification of $^{100}$Sr in ``Half-lives of the new isotopes $^{100}$Rb, $^{100}$Sr, $^{148}$Cs and of $^{199}$Rb, $^{99}$Sr and $^{147}$Cs'' in 1978 \cite{1978Kog01}. $^{100}$Sr was produced and identified by neutron induced fission of $^{235}$U at the On-line Separator f\"{u}r Thermisch Ionisierbare Spaltprodukte (OSTIS) facility in Grenoble, France. ``An improvement of the ion source of the online fission product separator OSTIS allowed us to identify the new isotopes $^{100}$Rb (50$\pm$10~msec), $^{100}$Sr (170$\pm$80~ms), and $^{148}$Cs (130$\pm$40~ms).'' This half-life of $^{100}$Sr is included in the weighted average of the accepted value of 202(3)~ms.

\subsection{$^{101}$Sr}\vspace{0.0cm}

In the 1983 paper ``Rotational structure and Nilsson orbitals for highly deformed odd-A nuclei in the A$\sim$ 100 region'' Wohn et al. described the discovery of $^{101}$Sr \cite{1983Woh01}. $^{101}$Sr was produced by neutron irradiation of $^{235}$U at the Brookhaven high-flux reactor and identified with the on-line mass separator TRISTAN. ``Using a high-temperature surface-ionization ion source, we have studied decays of $^{99}$Sr, $^{101}$Sr, $^{99}$Rb, and $^{101}$Y and found half-lives (in milliseconds) of 266$\pm$5, 121$\pm$6, 52$\pm$5, and 500$\pm$50, respectively.'' The half-life reported for $^{101}$Sr is included in the weighted average of the accepted half-life of 118(3)~ms.

\subsection{$^{102}$Sr}\vspace{0.0cm}

The first observation of $^{102}$Sr was documented by Hill et al. in the 1986 paper ``Identification and decay of neutron-rich $^{102}$Sr and level structure of A$\sim$100 Y nuclei'' \cite{1986Hil01}. $^{101}$Sr was produced by neutron irradiation of $^{235}$U at the Brookhaven high-flux reactor and identified with the on-line mass separator TRISTAN. ``The previously unreported decay of $^{102}$Sr to levels in $^{102}$Y has been studied from mass-separated activity produced in the the thermal neutron fission of $^{235}$U. A half-life for $^{102}$Sr was measured to be 68$\pm$8~ms.'' This half-life is included in the weighted average of the accepted value of 69$\pm$6 ms.

\subsection{$^{103-105}$Sr}\vspace{0.0cm}

$^{103}$Sr, $^{104}$Sr and $^{105}$Sr were discovered by Bernas et al. in 1997, as reported in ``Discovery and cross-section measurement of 58 new fission products in projectile-fission of 750$\cdot$AMeV $^{238}$U'' \cite{1997Ber01}. The experiment was performed using projectile fission of $^{238}$U at 750~MeV/nucleon on a beryllium target at GSI in Germany. ``Fission fragments were separated using the fragment separator FRS tuned in an achromatic mode and identified by event-by-event measurements of $\Delta$E-B$\rho$-ToF and trajectory.'' During the experiment, individual counts for $^{103}$Sr (409), $^{104}$Sr (72), and $^{105}$Sr (8) were recorded.

\subsection{$^{106,107}$Sr}\vspace{0.0cm}

The discovery of $^{106}$Sr and $^{107}$Sr was reported in the 2010 article ``Identification of 45 new neutron-rich isotopes produced by in-flight fission of a $^{238}$U Beam at 345 MeV/nucleon,'' by Ohnishi et al. \cite{2010Ohn01}. The experiment was performed at the RI Beam Factory at RIKEN, where the new isotopes were created by in-flight fission of a 345 MeV/nucleon $^{238}$U beam on a beryllium target. $^{106}$Sr and $^{107}$Sr were separated and identified with the BigRIPS superconducting in-flight separator. The results for the new isotopes discovered in this study were summarized in a table. Twenty-two counts for $^{106}$Sr and two counts for $^{107}$Sr were recorded.

\section{Discovery of $^{83-117}$Mo}

Thirty-five molybdenum isotopes from A = 83--117 have been discovered so far; these include 7 stable, 10 neutron-deficient and 18 neutron-rich isotopes. According to the HFB-14 model \cite{2007Gor01}, $^{133}$Mo should be the last odd-even particle stable neutron-rich nucleus while the even-even particle stable neutron-rich nuclei should continue through $^{146}$Mo. At the proton dripline four more isotopes $^{79-82}$Mo should be particle stable and in addition $^{76-78}$Mo could have half-lives longer than 10$^{-21}$~s \cite{2004Tho01}. Thus, about 30 isotopes have yet to be discovered corresponding to 46\% of all possible molybdenum isotopes.

\begin{figure}
	\centering
	\includegraphics[scale=.5]{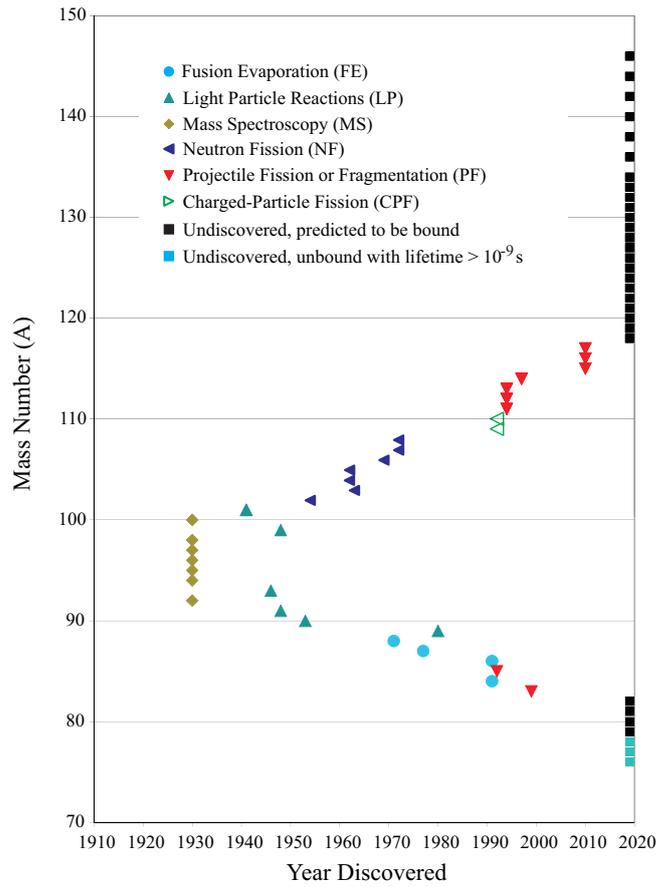}
	\caption{Molybdenum isotopes as a function of time when they were discovered. The different production methods are indicated. The solid black squares on the right hand side of the plot are isotopes predicted to be bound by the HFB-14 model. On the proton-rich side the light blue squares correspond to unbound isotopes predicted to have lifetimes larger than $\sim 10^{-9}$~s.}
\label{f:year-molybdenum}
\end{figure}

Figure \ref{f:year-molybdenum} summarizes the year of first discovery for all molybdenum isotopes identified by the method of discovery. The range of isotopes predicted to exist is indicated on the right side of the figure. The radioactive molybdenum isotopes were produced using fusion-evaporation reactions (FE), light-particle reactions (LP), neutron (NF) and charged-particle induced (CPF) fission, and projectile fragmentation or projectile fission (PF). The stable isotopes were identified using mass spectroscopy (MS). Light particles also include neutrons produced by accelerators. The discovery of each molybdenum isotope is discussed in detail and a summary is presented in Table 1.

\subsection{$^{83}$Mo}\vspace{0.0cm}

$^{83}$Mo was first reported in ``Observation of the Z = N + 1 nuclei $_{39}^{77}$Y, $_{40}^{79}$Zr, and $_{42}^{83}$Mo'' in 1999 by Janas et al. \cite{1999Jan01}. At GANIL, France, nickel targets were bombarded with a 60 MeV/nucleon $^{92}$Mo beam. $^{83}$Mo was separated with the LISE3 spectrometer and the kinetic energy, energy loss, and time-of-flight were measured. ``The projections of the T$_{z}$ = 0 and $-1/2$ species onto the Z axis are presented in [the figure] and clearly show the presence of the even-Z, Z = N + 1 nuclei, $^{75}_{38}$Sr, $^{79}_{40}$Zr, and $^{83}_{42}$Mo in our spectra.''

\subsection{$^{84}$Mo}\vspace{0.0cm}

The 1991 article ``The observation of $^{84}$Mo'' by Gelletly et al. described the discovery of $^{84}$Mo \cite{1991Gel01}. An enriched $^{28}$Si target was bombarded with a 195 MeV $^{58}$Ni beam at the Daresbury Nuclear Structure Facility and $^{84}$Mo was formed in the fusion-evaporation reaction $^{28}$Si($^{58}$Ni,2n). The reaction products were separated with the Daresbury Recoil Separator and identified by measuring $\gamma$-rays with 20 Compton suppressed Ge detectors. ``A 443.8$\pm$0.3~keV gamma ray was observed from $^{84}$Mo which is interpreted as the 2$^+-$0$^+$ transition in this nucleus.''

\subsection{$^{85}$Mo}\vspace{0.0cm}

The discovery of $^{85}$Mo is credited to Yennello et al. with their 1992 paper ``New nuclei along the proton-drip line near Z = 40,'' \cite{1992Yen01}.  At the National Superconducting Cyclotron Laboratory at Michigan State University, a 70~MeV/A $^{92}$Mo beam was produced by the K1200 cyclotron and impinged on a $^{58}$Ni target. $^{85}$Mo was identified with the A1200 fragment analyzer by measuring the time-of-flight and energy loss of the fragments. ``The mass spectra for residues with Z from 39 to 44 are shown in [the figure] with the new isotopes marked by arrows. Although both $^{84}$Mo and $^{86}$Mo have been previously observed, no reference to the identification of $^{85}$Mo was found. The other new isotopes observed in this study are $^{78}$Y, $^{82}$Nb, $^{86}$Tc, and $^{89,90}$Ru.''

\subsection{$^{86}$Mo}\vspace{0.0cm}

``Systematic behavior of the neutron-deficient molybdenum nuclei'' was published in 1991 by Gross et al. documenting the observation of $^{86}$Mo \cite{1991Gro01}. $^{86}$Mo was formed in the fusion-evaporation reactions $^{40}$Ca($^{50}$Cr,p2n) and $^{58}$Ni($^{32}$S,2p2n) at beam energies of 170 and 110 MeV, respectively. Reaction products were separated with the Daresbury Recoil Separator and identified with the POLYTESSA array. ``Despite the contamination of the 568-keV $^{86}$Mo transition by the 566-keV $\gamma$ ray from $^{86}$Zr, unambiguous identification of two transitions in $^{86}$Mo were possible.''

\subsection{$^{87}$Mo}\vspace{0.0cm}

``Investigation of neutron deficient Zr and Nb nuclei with heavy ion induced compound reactions,'' published in 1977 by Korschinek et al., described the first observation of $^{87}$Mo \cite{1977Kor01}. $^{32}$S beams with energies between 93 and 120 MeV were used to bombard enriched $^{58}$Ni targets at the Munich MP tandem accelerator. $^{87}$Mo was formed in the fusion evaporation reaction $^{58}$Ni($^{32}$S,n2p) and identified by the $\gamma$-ray spectra measured with a set of coaxial Ge(Li) detectors. ``We observe a growth with T$_{1/2}$ = 14.6$\pm$1.5~s for all strong activity lines from the decay $^{87}$Nb$\rightarrow^{87}$Zr except for the 134.9 keV line. This growth can only be explained by the existence of a 14.6~s state located either in $^{87}$Nb or $^{87}$Mo, thus feeding the 2.6~min state in $^{87}$Nb. Since it would be difficult with respect to the observed spins  of levels in the neighbouring nuclei to explain three long lived states in $^{87}$Nb the assignment of the 14.6~s activity to $^{87}$Mo seems most probable to us.'' The half-life of 14.6(15)~s is included in the weighted average for the presently accepted value of 14.02(26)~s.

\subsection{$^{88}$Mo}\vspace{0.0cm}

In ``Decay scheme studies of short-lived isotopes of 69 $\leqq$ A $\leqq$ 88 produced by heavy-ion bombardment,'' published in 1971, Doron and Blann described the first observation of $^{88}$Mo \cite{1971Dor01}. Enriched $^{58}$Ni and cobalt targets were bombarded with 85 to 100 MeV $^{32}$S beams at the University of Rochester MP Tandem Van de Graaff accelerator. $^{88}$Mo was identified by measuring half-lives, excitation functions and $\gamma$-rays. ``Three $\gamma$-rays decaying with a half-life of 8.2(5)~min were observed in spectra obtained by the irradiation of $^{59}$Co and $^{58}$Ni with a $^{32}$S beam. These $\gamma$-rays are listed in [a table]. The shape of the excitation function for this activity as obtained from the $^{59}$Co + $^{32}$S bombardments, is in excellent agreement with the theoretical predictions for the formation of $^{88}$Mo as may be seen by comparing [the figures].'' This half-life agree with the presently adopted value of 8.0(2)~min. A previously reported 27.3(14)~min half-life was evidently incorrect \cite{1963But01}.

\subsection{$^{89}$Mo}\vspace{0.0cm}

Pardo et al. observed $^{89}$Mo in 1980 as reported in ``Mass and excited states of the nucleus $^{89}$Mo'' \cite{1980Par01}. A molybdenum target was bombarded with a 70~MeV $^{3}$He beam produced by the Michigan State University cyclotron and $^{89}$Mo was formed in the reaction $^{92}$Mo($^3$He,$^6$He)$^{89}$Mo. $^{89}$Mo was identified in the focal plane of an Enge split-pole spectrograph with a two wire, charge-division proportional counter. ``The mass excess of $^{89}$Mo has been measured using the $^{92}$Mo($^3$He,$^6$He) reaction at 70-MeV bombarding energy. The mass excess was determined to be $-$75.008$\pm$0.015 MeV.'' A previously reported 7.1~min half-life for $^{89}$Mo was evidently incorrect \cite{1963But01} and another attempt to observe $^{89}$Mo was unsuccessful \cite{1975Hag01}.

\subsection{$^{90}$Mo}\vspace{0.0cm}

Diamond announced the discovery of $^{90}$Mo in ``Molybdenum 90'' in 1953 \cite{1953Dia01}. The Harvard University synchrocyclotron accelerated protons to 55-60MeV which then bombarded niobium metal foils. The resulting activity was measured with a Geiger counter following chemical separation. ``In summary, it may be said that from this work Mo$^{90}$ appears to have (1) a half-life of 5.7$\pm$0.2 hours; (2) a disintegration scheme involving predominantly three gamma-rays, with energies of approximately 1.1, 0.24-0.26, 0.10-0.13 Mev, of which the second gamma-ray is electron converted to a small extent, and the third to a much larger degree; (3) positrons of roughly 1.4-MeV maximum energy (or of an energy slightly greater than the maximum energy of those from Nb$^{90}$, and a greater amount of electron capture relative to positron emission than is the case with Nb$^{90}$.'' This half-life agrees with the currently adopted value of 5.56(9)~h.

\subsection{$^{91}$Mo}\vspace{0.0cm}

Sagane et al. reported the first observation of $^{91}$Mo in the 1938 paper ``A preliminary report on the radioactivity produced in Y, Zr, and Mo'' \cite{1938Sag02}. Fast neutrons produced by 3 MeV deuterons on lithium and beryllium at the cyclotron of the Institute of Physical and Chemical Research in Tokyo, Japan, irradiated zirconium targets. Positrons were detected following chemical separation. No further details were given and a half-life of 17~min was listed for $^{91}$Mo in a table. This value is close to the accepted half-life of 15.49(1)~min. Molybdenum activities of 17~min \cite{1937Bot03} and 21~min \cite{1937Hey01} were reported earlier without mass assignments.

\subsection{$^{92}$Mo}\vspace{0.0cm}

The discovery of stable $^{92}$Mo was published in 1930 by Aston in ``Constitution of molybdenum'' \cite{1930Ast04}. Molybdenum carbonyl was used in the Cavendish mass spectrograph and no further experimental details were given. ``The following are the mass numbers and their approximate percentage abundance: 92 (14.2), 94 (10.0), 95 (15.5), 96 (17.8), 97 (9.6), 98 (23.0) 100 (9.8).''

\subsection{$^{93}$Mo}\vspace{0.0cm}

Wiedenbeck described the observation of $^{93}$Mo in the 1946 paper ``Radioactive isotopes in the columbium region''. Columbium (niobium) metal foils were bombarded with 10 MeV deuterons. Beta and $\gamma$-ray spectra as well as absorption curves were measured. Short bombardments of columbium metal foils with 10-Mev deuterons produced the 6.6-minute $\beta^-$-activity in Cb$^{94}$ as well as 18-minute and 6.5-hour positron activities. The latter periods can be assigned to isomeric states of Mo$^{93}$ produced in the Cb$^{93}$(d,2n) process.'' This half-life agrees with the currently adopted value of 6.85(7)~h for the isomeric state.

\subsection{$^{94-98}$Mo}\vspace{0.0cm}

The discovery of stable $^{94}$Mo, $^{95}$Mo, $^{96}$Mo, $^{97}$Mo, and $^{98}$Mo was published in 1930 by Aston in ``Constitution of molybdenum'' \cite{1930Ast04}. Molybdenum carbonyl was used in the Cavendish mass spectrograph and no further experimental details were given. ``The following are the mass numbers and their approximate percentage abundance: 92 (14.2), 94 (10.0), 95 (15.5), 96 (17.8), 97 (9.6), 98 (23.0) 100 (9.8).''

\subsection{$^{99}$Mo}\vspace{0.0cm}

Sagane et al. reported the first observation of $^{99}$Mo in the 1938 paper ``A preliminary report on the radioactivity produced in Y, Zr, and Mo'' \cite{1938Sag02}. Fast and slow neutrons produced by 3 MeV deuterons on lithium and beryllium at the cyclotron of the Institute of Physical and Chemical Research in Tokyo, Japan, irradiated zirconium targets. Positrons were detected following chemical separation. No further details were given and a half-life of 64~h was listed for $^{99}$Mo in a table. This value is close to the accepted half-life of 65.94(1)~h.

\subsection{$^{100}$Mo}\vspace{0.0cm}

The discovery of stable $^{100}$Mo was published in 1930 by Aston in ``Constitution of molybdenum'' \cite{1930Ast04}. Molybdenum carbonyl was used in the Cavendish mass spectrograph and no further experimental details were given. ``The following are the mass numbers and their approximate percentage abundance: 92 (14.2), 94 (10.0), 95 (15.5), 96 (17.8), 97 (9.6), 98 (23.0) 100 (9.8).''

\subsection{$^{101}$Mo}\vspace{0.0cm}

The discovery of $^{101}$Mo is credited to the 1941 paper ``Untersuchung \"uber das `19-Minuten'-Isotop von Molybd\"an und das daraus entstehende Isotop von Element 43'' by Maurer and Ramm \cite{1941Mau01}. Slow neutrons from the bombardment of beryllium with deuterons were used to activate molybdenum targets. The resulting activities were measured following chemical separation. ``Auf Grund der von uns gemessenen fast gleichen Halbwertszeiten von 14,6 Minuten f\"ur Molybd\"an und 14,0 Minuten f\"ur Element 43, lassen sich obige Widerspr\"uche als nur scheinbare jedoch leicht auffkl\"aren.'' [Based on our measurements of almost identical half-lives of 14.6~min for molybdenum and 14.0 for element 43, the above mentioned apparent contradiction can easily be resolved.] This half-life agrees with the presently accepted value of 14.61(3)~min. Hahn and Strassmann simultaneously submitted two papers reporting the same half-life for $^{101}$Tc \cite{1941Hah01,1941Hah03} giving Maurer and Ramm credit for the solution to the puzzle of the equal half-lives of $^{101}$Mo and $^{101}$Tc.

\subsection{$^{102}$Mo}\vspace{0.0cm}

$^{102}$Mo was discovered by Wiles and Coryell as described in the 1954 paper ``Fission yield fine structure in the mass region 99-106'' \cite{1954Wil01}. Thermalized neutrons from the Be(d,n) reaction produced with 15 MeV deuterons from the M.I.T. cyclotron irradiated uranium targets. Beta-decay curves were recorded following chemical separation. $^{102}$Mo was also produced in photo-fission and deuteron induced fission of enriched U$^{238}$. ``Fission yields have been determined for 14.6-min Mo$^{101}$ and 11.0-min Mo$^{102}$ relative to 67-hr Mo$^{99}$, and for 67-hr Mo$^{99}$, 43-day Ru$^{103}$, 4.5-hr Rh$^{105}$, and 1.0-yr Ru$^{106}$, all relative to 12.8-day Ba$^{140}$.'' This half-life for $^{102}$Mo of 11~min agrees with the currently adopted value of 11.3(2)~min. Wiles and Coryell did not consider this the discovery of $^{102}$Mo referring to a paper by themselves listed as `to be published'. Previously a half-life of 12~min was reported without a mass assignment  \cite{1941Hah03,1941Hah01}. Less than a month after the submission by Wiles and Coryell, Flegenheimer and Seelmann-Eggebert assigned the 11.5(5)~min half-life tentatively to $^{102}$Mo \cite{1954Fle01}.

\subsection{$^{103}$Mo}\vspace{0.0cm}

$^{103}$Mo was identified in 1963 by Kienle et al. in the paper ``Nachweis kurzlebiger Molybd\"an-Spaltprodukte und Zerfall ihrer Technetium-T\"ochter III. Zerfallsreihe: $^{103}$Mo$-^{103}$Tc$-^{103}$Ru'' \cite{1963Kie01}. Uranium oxide and uranyl acetate were bombarded with neutrons and the fission products were chemically separated. $^{103}$Mo was identified by measuring $\gamma$-ray spectra following chemical separation. ``Die Halbwertszeit des bisher unbekannten $^{103}$Mo, der Mutter von $^{103}$Tc, wurde aus dem Verh\"altnis der $^{103}$Tc (T$_{1/2}$ = 50 sec) zur $^{104}$Tc (T$_{1/2}$ = 18 min) Aktivit\"at als Funktion der Abtrennzeit der Molybd\"anisotope von den Spaltprodukten bestimmt.'' [The half-life of the hitherto unknown $^{103}$Mo - the mother of $^{103}$Tc - was determined from the $^{103}$Tc (T$_{1/2}$ = 50 sec) to $^{104}$Tc (T$_{1/2}$ = 18 min) activity ratio as function of the separation time of molybdenum isotopes from the fission products.] The reported half-life of 70(10)~s agrees with the presently adopted value of 67.5(15)~s.

\subsection{$^{104}$Mo}\vspace{0.0cm}

The discovery of $^{104}$Mo was reported in the article ``Nachweis Kurzlebiger Molybd\"anspaltisotope und Zerfall ihrer Technetium-Tochter. I. Zerfallsreihe: Mo$^{104}-$Tc$^{104}$'' published in 1962 by Kienle et al. \cite{1962Kie01}. Uranium fission products were chemically separated and $\beta$-activities and $\gamma$-spectra measured. ``[Figure] zeigt A($^{104}$Tc/$^{101}$Tc) als Funktion der Wartezeit t$_V$. Die einzelnen Me\ss punkte sind bereits auf die Halbwertszeit von Mo$^{101}$ korrigiert. Daraus ergibt sich f\"ur den Zerfall des Spaltisotopes Mo$^{104}$ eine Halbwertszeit von 1.1$\pm$0.1~min.[[The figure] shows A($^{104}$Tc/$^{101}$Tc) as a function of the waiting period t$_V$. The individual data points are already corrected for the half-life of $^{101}$Mo. This results in a half-life of 1.1$\pm0.1$~min for the decay of the fission product $^{104}$Mo.] This half-life agrees with the currently adopted value of 60(2)~s. A previously reported half-life of $\sim$60~d \cite{1947See01} was evidently incorrect.

\subsection{$^{105}$Mo}\vspace{0.0cm}

$^{105}$Mo was first observed by Kienle et al. in the 1962 article `Nachweis Kurzlebiger Molybd\"anspaltisotope und Zerfall ihrer Technetium-Tochter. II. Zerfallsreihe: Mo$^{105}-$Tc$^{105}-$Ru$^{105}$'' \cite{1962Kie02}. Chrome-hexacarbonyl and uranylacetat were irradiated with neutrons and the subsequent $\beta$- and $\gamma$-ray activities were measured following chemical separation. ``In [der Figur] ist die Aktivit\"at von Tc$^{105}$, bezogen auf die Aktivit\"at von Tc$^{101}$ nach verschiedenen Wartezeiten bis zur Molybd\"anabtrennung aufgetragen. Daraus ergibt sich nach einer Korrektur auf die Halbwertszeit von Mo$^{101}$ die Halbwertszeit von Mo$^{105}$ zu 40 sec.'' [The activity of Tc$^{105}$ relative to the Tc$^{101}$ is plotted for different waiting periods of the molybdenum separation. This results in a half-life for Mo$^{105}$ of 40~sec after the correction for the half-life of Mo$^{101}$.] This is consistent with the presently accepted value of 35.6(16)~s. Earlier attempts to observe $^{105}$Mo could only extract an upper limit for the half-life of 2~min \cite{1955Fle01}. A previously reported half-life of $\sim$5~min \cite{1947See01} was evidently incorrect.

\subsection{$^{106}$Mo}\vspace{0.0cm}

Hastings et al. discovered $^{106}$Mo as reported in the 1969 paper ``Fractional cumulative yields of $^{103}$Mo, $^{105}$Mo and $^{106}$Mo from thermal-neutron induced fission of $^{235}$U'' \cite{1969Has01}. Enriched $^{235}$U was irradiated with thermal neutrons in the pneumatic tube facility of the Oak Ridge Research Reactor. The resulting activities were measured with a sodium iodide scintillation crystal and a Geiger-M\"uller detector following chemical separation. ``Since no half-life has been reported for $^{106}$Mo, results were obtained from a least-squares analysis in which both intercept and half-life were allowed to vary. A value of 9.5$\pm$0.5 sec was obtained for the half-life of $^{106}$Mo.'' This half-life agrees with the presently adopted value of 8.73(12)~s. In an earlier measurement, only an upper limit of 10~s was determined \cite{1965Von01}.

\subsection{$^{107,108}$Mo}\vspace{0.0cm}

$^{107}$Mo and $^{108}$Mo were first identified by Trautmann et al. in the 1972 paper ``Identification of short-lived isotopes of zirconium, niobium, molybdenum, and technetium in fission by rapid solvent extraction techniques'' \cite{1972Tra01}. $^{235}$U and $^{239}$Pu targets were irradiated with thermal neutrons at the Mainz Triga reactor. Following chemical separation, $\gamma$-ray spectra were recorded with a Ge(Li) detector. ``For $^{107}$Mo, only an estimation of the half-life, about 5 sec, was possible since the growth curve shows a complex behaviour. The half-life of $^{108}$Mo, 1.5$\pm$0.4 sec, agrees within the errors with a value of 0.86$\pm$0.39 sec recently reported by Wilhelmy et al.'' These measured values are consistent with the presently adopted values of 3.5(5)~s and 1.09(2)~s for $^{107}$Mo and $^{108}$Mo, respectively. The accepted half-life for $^{107}$Mo corresponds to a later measurement by the same group \cite{1977Tit01}. The $^{108}$Mo reference was an internal report \cite{1970Wil02}.

\subsection{$^{109,110}$Mo}\vspace{0.0cm}

The first observation of $^{109}$Mo and $^{110}$Mo was reported by \"Ayst\"o et al. in ``Discovery of rare neutron-rich Zr, Nb, Mo, Tc and Ru isotopes in fission: Test of $\beta$ half-life predictions very far from stability'' in 1992 \cite{1992Ays01}. At the Ion Guide Isotope Separator On-Line (IGISOL) in Jyv\"askyl\"a, Finland, targets of uranium were bombarded with 20 MeV protons. $\beta$ decays were measured with a planar Ge detector, while $\gamma$-rays were measured with a 50\% Ge detector located behind a thin plastic detector. ``The data show clearly K$\alpha$ peaks associated with $\beta$ decay of the new isotopes $^{107}$Nb, $^{109}$Mo, $^{110}$Mo, and $^{113}$Tc.'' The measured half-life of 520(60)~ms corresponds to the currently accepted value for $^{109}$Mo and the half-life of 250(100)~ms for $^{110}$Mo agrees with the presently adopted value of 300(40)~ms.

\subsection{$^{111-113}$Mo}\vspace{0.0cm}

Bernas et al. discovered $^{111}$Mo, $^{112}$Mo, and $^{113}$Mo in 1994 at GSI, Germany, as reported in ``Projectile fission at relativistic velocities: A novel and powerful source of neutron-rich isotopes well suited for in-flight isotopic separation'' \cite{1994Ber01}. The isotopes were produced using projectile fission of $^{238}$U at 750 MeV/nucleon on a lead target. ``Forward emitted fragments from $^{80}$Zn up to $^{155}$Ce were analyzed with the Fragment Separator (FRS) and unambiguously identified by their energy-loss and time-of-flight.'' The experiment yielded 135, 29 and 3 individual counts of $^{111}$Mo, $^{112}$Mo, and $^{113}$Mo, respectively.

\subsection{$^{114}$Mo}\vspace{0.0cm}

$^{114}$Mo was discovered by Bernas et al. in 1997, as reported in ``Discovery and cross-section measurement of 58 new fission products in projectile-fission of 750$\cdot$AMeV $^{238}$U'' \cite{1997Ber01}. The experiment was performed using projectile fission of $^{238}$U at 750~MeV/nucleon on a beryllium target at GSI in Germany. ``Fission fragments were separated using the fragment separator FRS tuned in an achromatic mode and identified by event-by-event measurements of $\Delta$E-B$\rho$-ToF and trajectory.'' During the experiment, eight counts for $^{114}$Mo were recorded.

\subsection{$^{115-117}$Mo}\vspace{0.0cm}

The discovery of $^{115}$Mo, $^{116}$Mo, and $^{117}$Mo was reported in the 2010 article ``Identification of 45 new neutron-rich isotopes produced by in-flight fission of a $^{238}$U beam at 345 MeV/nucleon,'' by Ohnishi et al. \cite{2010Ohn01}. The experiment was performed at the RI Beam Factory at RIKEN, where the new isotopes were created by in-flight fission of a 345 MeV/nucleon $^{238}$U beam on a beryllium target. The isotopes were separated and identified with the BigRIPS superconducting in-flight separator. The results for the new isotopes discovered in this study were summarized in a table. 993 counts for $^{115}$Mo, 78 counts for $^{116}$Mo, and 6 counts for $^{117}$Mo were recorded.

\section{Discovery of $^{89-126}$Rh}

Thirty-eight rhodium isotopes from A = 89--126 have been discovered so far; these include 1 stable, 14 neutron-deficient and 23 neutron-rich isotopes. According to the HFB-14 model \cite{2007Gor01}, $^{142}$Rh should be the last odd-odd particle stable neutron-rich nucleus while the odd-even particle stable neutron-rich nuclei should continue through $^{151}$Rh. The proton dripline has most likely been reached at $^{89}$Rh but four isotopes, $^{85-88}$Rh could have half-lives longer than 10$^{-21}$~s \cite{2004Tho01}. Thus, about 25 isotopes have yet to be discovered corresponding to 40\% of all possible rhodium isotopes.

\begin{figure}
	\centering
	\includegraphics[scale=.5]{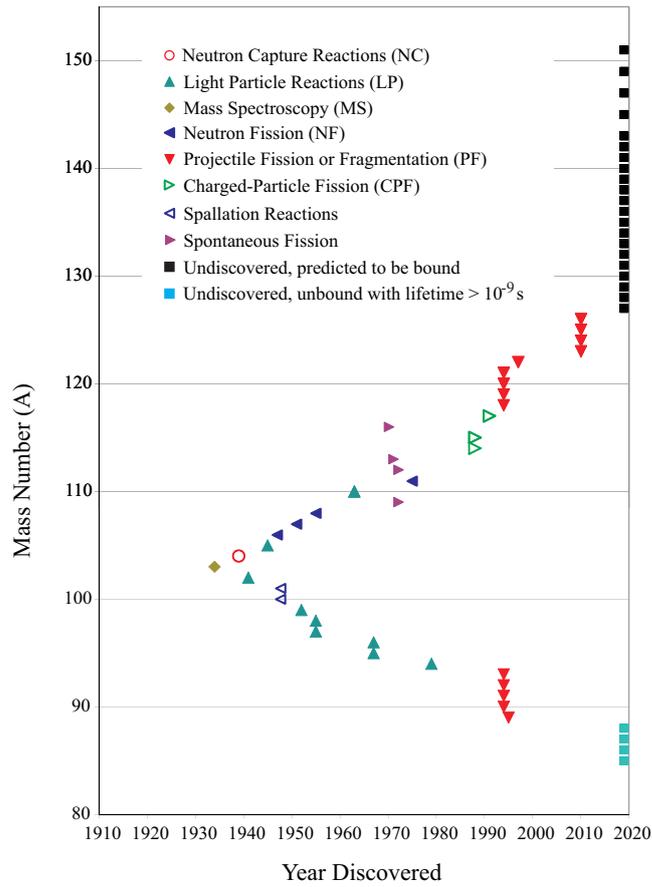}
	\caption{Rhodium isotopes as a function of time when they were discovered. The different production methods are indicated. The solid black squares on the right hand side of the plot are isotopes predicted to be bound by the HFB-14 model. On the proton-rich side the light blue squares correspond to unbound isotopes predicted to have lifetimes larger than $\sim 10^{-9}$~s.}
\label{f:year-rhodium}
\end{figure}

Figure \ref{f:year-rhodium} summarizes the year of first discovery for all rhodium isotopes identified by the method of discovery. The range of isotopes predicted to exist is indicated on the right side of the figure. The radioactive rhodium isotopes were produced using neutron capture (NC), light-particle reactions (LP), spontaneous (SF), neutron induced (NF) and charged-particle induced (CPF) fission, spallation reactions (SP), and projectile fragmentation or projectile fission (PF). The stable isotopes were identified using mass spectroscopy (MS). Light particles also include neutrons produced by accelerators. The discovery of each rhodium isotope is discussed in detail and a summary is presented in Table 1.

\subsection{$^{89}$Rh}\vspace{0.0cm}

Rykaczewski et al. discovered $^{89}$Rh in their 1995 paper ``Identification of new nuclei at and beyond the proton drip line near the doubly magic nucleus $^{100}$Sn'' \cite{1995Ryk01}. A 63 MeV/nucleon $^{112}$Sn beam from the GANIL cyclotron complex bombarded a natural nickel target. $^{89}$Rh was identified with the Alpha and LISE3 spectrometers. ``The obtained data have allowed also for the identification of six other new nuclei, namely $^{103}$Sb, $^{104}$Sb, $^{98}$In, $^{91}$Pd, $^{89}$Rh, and $^{87}$Ru, which are clearly isolated from the neighboring heavier isotopes in the mass spectra of [the figure].''

\subsection{$^{90-93}$Rh}\vspace{0.0cm}

In ``Identification of new nuclei near the proton drip line,'' Hencheck et al. reported the discovery of $^{90}$Rh, $^{91}$Rh, $^{92}$Rh, and $^{93}$Rh in 1994 \cite{1994Hen01}. A $^{106}$Cd beam accelerated to 60 MeV/u at the National Superconducting Cyclotron Laboratory (NSCL) at Michigan State University bombarded a natural nickel target. The isotopes were analyzed with the A1200 projectile fragment separator and identified event-by-event with measurements of the magnetic rigidity, time of flight, energy-loss, and total energy. ``A number of new nuclides were identified including $^{88}$Ru, $^{90,91,92,93}$Rh, $^{92,93}$Pd, and $^{94,95}$Ag.''

\subsection{$^{94}$Rh}\vspace{0.0cm}

``The $\gamma$-spectrum of $^{94}$Rh'' published in 1979 by Zarifov et al. reported the observation of $^{94}$Rh \cite{1979Zar01}. Enriched $^{96}$Ru targets were bombarded with 34-59 MeV $^{3}$He from an isochronous cyclotron. $^{94}$Rh was produced in the reaction $^{96}$Ru($^3$He,p4n) and identified by measuring characteristic $\gamma$-rays with a Ge(Li) detector. ``Along with the $\gamma$-lines of known radio-nuclides unknown $\gamma$-lines having energies and relative intensities (indicated in parentheses) of 311.1 (90$\pm$10), 756.5 (70$\pm$20), and 1430.4 (100) keV were present in the $\gamma$-spectra. Based on the identification given above, these $\gamma$ lines were assigned to the decay of $^{94}$Rh. According to our measurements the half-life of $^{94}$Rh is equal to 24$\pm$3 sec.'' This half-life agrees with the presently adopted value of 25.8(2)s for the isomeric state. A half-life of 25(3)~s had previously been reported in a conference abstract \cite{1973Wei01}.

\subsection{$^{95,96}$Rh}\vspace{0.0cm}

In the 1967 paper ``Rhodium-96 and Rhodium-95'' Aten et al. identified $^{95}$Rh and $^{96}$Rh \cite{1967Ate01}. Protons irradiated enriched $^{96}$Ru and the two isotopes $^{95}$Rh and $^{96}$Rh were produced in (p,2n) and (p,n) reactions, respectively. The isotopes were identified after chemical separation by measuring annihilation radiation and $\gamma$-ray spectra. ``On the basis of these arguments we have recognized two isomers of $^{96}$Rh, one with a half-life of 1.5$\pm$0.5 minutes, and the other one with 9.25$\pm$1 minutes... Above 17 MeV another rhodium activity appears, which becomes progressively stronger compared to $^{96}$Rh at higher energies. This nuclide which we identify as $^{95}$Rh, has a half-life of 4.75($\pm$0.5) minutes.'' The latter half-life agrees with the presently adopted value of 5.02(10)~min for $^{95}$Rh. The measured half-lives for $^{96}$Rh agrees with the values of 9.90(10)~min and 1.51(2)~min for the ground state and the isomeric state, respectively.

\subsection{$^{97,98}$Rh}\vspace{0.0cm}

``Formation and properties of neutron-deficient isotopes of rhodium and palladium'' by Aten Jr and De Vries-Hamerling describes the identification of $^{97}$Rh and $^{98}$Rh in 1955 \cite{1955Ate01}. Ruthenium targets were bombarded with 17 MeV and 25 MeV deuterons from the Amsterdam Instituut voor Kernphysisch Onderzoek Phillips cyclotron. Beta- and $\gamma$-ray spectra were measured following chemical separation. For the discovery of $^{97}$Rh they stated: ``In an earlier paper from this institute, it was suggested that the 32-minutes rhodium isotope, which was first observed by Eggen and Pool, might be $^{97}$Rh. In this case $^{97}$Ru should grow from it. We have since been able to isolate $^{97}$Ru from the rhodium fraction obtained by irradiating ruthenium with deutons... The half-life of the rhodiummother we calculated to be 35$\pm$10 minutes.'' In the earlier paper mentioned in the quote a half-life of 31~min was measured but a firm mass assignment was not possible \cite{1952Ate01}. This half-life agrees with the presently accepted value of 30.7(6)~min. Previously a 32~min half-life was reported in a conference abstract without a mass assignment \cite{1949Egg01}.

The identification of $^{98}$Rh was also a confirmation of an earlier uncertain mass assignment: ``When we observed that the palladium mother isotope of the 9-minute rhodium had a half-life of 15 minutes we suggested the mass number 98 and 96 for these nuclei. We have since found by means of milking experiments that the palladium-mother of the 9-minutes rhodium can also be obtained from ruthenium irradiated with 24 MeV helium ions, which rules out the mass number 96, as the latter can only be formed by an ($\alpha$,4n) process. Therefore it seems likely that the 9-minutes period is due to $^{98}$Rh and its 17-minutes mother to $^{98}$Pd.'' This half-life agrees with the presently adopted value of 3.6(2)~min for an isomeric state. The authors had discussed this half-life in two previous papers without a firm mass assignment \cite{1952Ate01,1953Ate01}.

\subsection{$^{99}$Rh}\vspace{0.0cm}

The earliest identification of $^{99}$Rh in the literature is the 1952 paper ``Radiations of Rh$^{99}$, Rh$^{101}$, Rh$^{105}$, and Ru$^{105}$'' by Scoville et al. \cite{1952Sco01}. Samples of high-purity ruthenium metal were bombarded with 6.3 MeV protons and the resulting activities were measured with a 180$^\circ$ spectrometer. ``The positron spectra of 4.5-hour Rh$^{99}$ and 19-hour Rh$^{100}$ were observed.'' This half-life is consistent with the presently adopted value of 4.7(1)~h for the isomeric state. Apparently, Scoville et al. assumed this activity to be known, however, the only previous publication was a conference abstract reporting a 5~h half-life with no mass assignment \cite{1949Egg01}.

\subsection{$^{100,101}$Rh}\vspace{0.0cm}

Lindner and Perlman discovered $^{100}$Rh and $^{101}$Rh in 1948 in ``Neutron deficient isotopes of tellurium and antimony'' \cite{1948Lin02}. A 200 MeV deuteron beam from the Berkeley 184-inch cyclotron bombarded an antimony target. Beta-decay curves as well as X- and $\gamma$-ray spectra were measured following chemical separation. For $^{100}$Rh, the authors stated: ``19.4-hr.Rh$^{100}$.$-$ This rhodium isotope could be removed from the palladium fraction which had been purified following the decay of 9-hr. Pd$^{101}$...'' and for $^{101}$Rh: ``4.3-day Rh$^{101}$.$-$ Rhodium removed from palladium contained only the 4.3-day Rh$^{101}$ after the 19.4-hr. Rh$^{100}$ had decayed.'' These half-lives agree with the currently adopted values of 20.8(1)~h and 4.34(1)~d for the ground state of $^{100}$Rh and the isomeric state of $^{101}$Rh, respectively.

\subsection{$^{102}$Rh}\vspace{0.0cm}

Minakawa published the discovery of  $^{102}$Rh in ``Long-lived activity of rhodium'' \cite{1941Min01}. A powdered metallic rhodium target was irradiated with fast neutrons produced in reactions with deuterons on lithium. The resulting activity was measured with a Lauritsen type electroscope and a thin-walled G-M counter after chemical separation. ``A long-lived activity decaying with a period of 210$\pm$6 days was found. Chemical separation showed that the carrier of this activity is rhodium... From the above observations it can be concluded that the corresponding isotope of this long-lived activity is Rh$^{102}$, produced from the abundant stable isotope Rh$^{103}$ by the reaction (n, 2n).'' This half-life agrees with the accepted value of 207.3(17)d.

\subsection{$^{103}$Rh}\vspace{0.0cm}

$^{103}$Rh was discovered by Aston in ``Constitution of hafnium and other elements'' in 1934 \cite{1934Ast02}. The stable isotope was identified with an anode discharge tube installed at the Cavendish Laboratory mass spectrograph. ``Rhodium gave the feeblest effect of any element yet analysed; only one line, that expected at 103, could be clearly detected.''

\subsection{$^{104}$Rh}\vspace{0.0cm}

The credit for the discovery of $^{104}$Rh is given to Crittenden for the 1939 paper ``The beta-ray spectra of Mg$^{27}$, Cu$^{62}$ and the nuclear isomers of Rh$^{104}$'' \cite{1939Cri01}. Rhodium samples were irradiated with neutrons produced by the bombardment of 1.2 MeV deuterons on lithium. The decay curves of the resulting activities were measured. ``The half-life of the long period activity was determined to be 4.37$\pm$0.05 minutes. Subtracting the ordinates of the line from the ordinates of the decay curve for small values of the time resulted in the curve shown in curve II [of the figure]. This gave a half-life of 42$\pm$2 seconds for the short period activity.'' These half-lives agree with the presently adopted values of 42.3(4)~s and 4.34(3)~min for the ground state and the isomeric state, respectively. Half-lives of 44~s \cite{1935Ama01,1938Pon02}, 3.9~min \cite{1935Ama01}, 4.2~min \cite{1938Pon02}, and 4.0~min \cite{1937Poo01} were previously reported without definite mass assignments. Crittenden did not consider his observation the discovery of $^{104}$Rn, giving credit to a paper by Pontecorvo \cite{1938Pon01}. However, Pontecorvo had assigned the observed activities incorrectly to $^{105}$Rh.

\subsection{$^{105}$Rh}\vspace{0.0cm}

$^{105}$Ru was discovered by Bohr and Hole in 1945, in their paper entitled ``Radioactivity induced by neutrons and deuterons in ruthenium'' \cite{1945Boh01}. Targets of natural ruthenium metal were bombarded with 5.5 MeV deuterons, as well as fast and slow neutrons, from the cyclotron at the Stockholm Forskningsinstitutet f\"or Fysik. The activities were measured with glass Geiger-Muller counters following chemical separation. ``We are thus forced to assign the two nuclei [4.4 h and 37 h] to $^{105}$Ru and $^{105}$Rh.'' The reported half-life of 37(1)~h agrees with the currently adopted value of 35.36(6)~h. Previously, half-lives of 39~h \cite{1936Liv01} and 34~h \cite{1941Nis01,1941Nis02} were reported without mass assignments. Also, 100~s \cite{1935Kou01}, and 45~d \cite{1938deV01} half-lives were assigned to $^{105}$Ru which was evidently incorrect. In addition, Pontecorvo had assigned the 44~s and 4.2~min half-lives of the ground state and isomeric state of $^{104}$Rh incorrectly to $^{105}$Rh.

\subsection{$^{106}$Rh}\vspace{0.0cm}

The first identification of $^{106}$Rh is credited to the 1947 paper ``Decay scheme of $_{45}$Rh$^{106}$'' by Peacock \cite{1947Pea01}. $^{106}$Rh was studied from a $^{106}$Ru sample chemically separated from other fission fragments. Beta- and $\gamma$-ray singles as well as $\beta$-$\gamma$ and $\gamma$-$\gamma$ coincidence spectra were measured. ``The decay scheme of $_{45}$Rh$^{106}$ (30s) is proposed as a complex beta-spectra with end points at 3.55$\pm$0.10 and 2.30$\pm$0.10 Mev.'' The 30~s activity had been reported as part of the Plutonium Project \cite{1951Gle02} without a mass assignment. It was independently observed by Grummitt and Wilkinson who also did not uniquely assign it to $^{106}$Ru; the mass number was only listed in brackets \cite{1948Gru01}. In addition, a 40~s half-life was previously reported also without a mass assignment \cite{1946See01}.

\subsection{$^{107}$Rh}\vspace{0.0cm}

Glendenin described the first observation of $^{107}$Rh in ``Short-lived ruthenium-rhodium decay chains,'' which was published in 1951 as part of the Plutonium Project Series \cite{1951Gle01}. Uranyl nitrate was irradiated with neutrons from the Clinton pile and $^{107}$Rh was identified by measuring its activity following chemical separation. ``The available mass numbers for the 4m Ru $-$ 24m Rh chain are thus limited to 107, 108, and 110 or greater... On the basis of energy a mass assignment of 107 is highly probable.'' The measured half-life agrees with the presently accepted value of 21.7(4)~min. This ruthenium - rhodium decay chain had been reported earlier without a mass assignment \cite{1943Bor03}.

\subsection{$^{108}$Rh}\vspace{0.0cm}

$^{108}$Rh was identified by Baro et al. in the 1955 paper ``Eine neue Isobarenreihe 108 (110)'' \cite{1955Bar01}. Uranyl was bombarded with 28 MeV deuterons, as well as fast and slow neutrons. Activities were measured with $\beta$-counters and $\gamma$-scintillation counters following chemical separation. ``Es ist daher wahrscheinlich da\ss\ die neue Isobarenreihe der Massenzahl 108 zugerechnet werden muss obwohl allerdings die Massenzahl 110 und h\"oher denkbar w\"aren.'' [It is thus probable that the new isobaric chain must be assigned to mass 108, although mass number 110 or higher could be possible.] The assigned half-life of 18(1)~s is consistent with the currently adopted value of 16.8(5)~s.

\subsection{$^{109}$Rh}\vspace{0.0cm}

In the 1972 paper ``Further measurements of gamma transitions in spontaneous-fission fragments of $^{252}$Cf,'' Hopkins et al. reported the observation of $^{109}$Rh \cite{1972Hop01}.  Fission fragments from the spontaneous fission of $^{252}$Cf were measured in coincidence with X-rays and low-energy $\gamma$-rays. In a table the energy (358.8~keV) of an excited state was identified correctly. This was later confirmed by Fettweis and del Marmol \cite{1975Fet01}.

\subsection{$^{110}$Rh}\vspace{0.0cm}

Karras and Kantele reported the discovery of $^{110}$Rh in the 1963 paper ``New nuclear species Sc$^{50m}$ and Rh$^{110}$'' \cite{1963Kar01}. A natural palladium and enriched $^{110}$Pd sample was irradiated with 14-15 MeV neutrons from the University of Arkansas neutron generator. The resulting activity was measured with two NaI(Tl) detectors. ``As expected, a fast decaying 375$\pm$5 keV gamma was found from an irradiated natural palladium metal sample, and since the intensity of this gamma was strongly increased when a 91.4\% enriched Pd$^{110}$ sample was bombarded, it is clear that the new activity must be due to a Pd$^{110}$+n reaction product. By following the 375 keV gamma photopeak area in many consecutive spectra the gamma was found to decay with a half-life of approximately 5 sec.'' This half-life is consistent with the presently adopted value of 3.2(2)s.

\subsection{$^{111}$Rh}\vspace{0.0cm}

The first identification of $^{111}$Rh was described by Franz and Herrmann in 1975 in ``Identification of short-lived neutron-rich ruthenium and rhodium isotopes in fission'' \cite{1975Fra02}. Thermal neutrons from the Mainz reactor irradiated $^{239}$Pu and $^{249}$Cf targets. Gamma-ray singles and coincidence spectra were recorded with two Ge(Li) detectors following chemical separation. ``We observed the growth and decay of a 275.3 keV $\gamma$-ray peak which we attribute to the 11$\pm$1 sec $^{111}$Rh daughter of $^{111}$Ru.'' This half-life corresponds to the currently accepted value which was reported by Franz and Herrmann three years later \cite{1978Fra01} confirming the first observation.

\subsection{$^{112}$Rh}\vspace{0.0cm}

In the 1972 paper ``Further measurements of gamma transitions in spontaneous-fission fragments of $^{252}$Cf,'' Hopkins et al. reported the observation of $^{112}$Rh \cite{1972Hop01}.  Fission fragments from the spontaneous fission of $^{252}$Cf were measured in coincidence with X-rays and low-energy $\gamma$-rays. In a table the energy (60.5~keV) of an excited state was identified. This value is mentioned in the ENSDF database but the level has not been placed in the level scheme \cite{2008ENS01}.

\subsection{$^{113}$Rh}\vspace{0.0cm}

In the 1971 paper ``Low-energy transitions from the deexcitation of spontaneous fission fragments of $^{252}$Cf,'' Hopkins et al. reported the observation of $^{113}$Rh \cite{1971Hop01}.  Fission fragments from the spontaneous fission of $^{252}$Cf were measured in coincidence with X-rays and low-energy $\gamma$-rays. In a table the energy (212.3~keV) of the first excited state was identified correctly. This energy had also been reported earlier \cite{1970Joh01}, however, no specific element assignment was made. This state (211.72~keV) is included in the level scheme accepted by ENSDF\cite{2008ENS01}.

\subsection{$^{114}$Rh}\vspace{0.0cm}

$^{114}$Rh was identified by \"Ayst\"o et al. in the 1988 paper ``Levels in $^{110}$Pd, $^{112}$Pd, $^{114}$Pd and $^{116}$Pd from the beta decays of the on-line mass separated Rh isotopes'' \cite{1988Ays02}. $^{114}$Rh were produced at the University of Jyv\"askyl\"a in Finland by proton-induced fission of $^{238}$U. The isotopes were separated with the IGISOL on-line isotope separator and identified by $\gamma$-ray, x-ray, and $\beta$-ray emissions. ``Two different half-lives were obvious for $^{112}$Rh, whereas only one could be seen for $^{114}$Rh.'' The measured half-life of 1.85(5)~s listed in a table corresponds to the presently accepted half-life of $^{114}$Rh.

\subsection{$^{115}$Rh}\vspace{0.0cm}

\"Ayst\"o et al. discovered $^{115}$Rh in ``Identification and decay of new neutron-rich isotopes $^{115}$Rh and $^{116}$Rh'' in 1988 \cite{1988Ays01}. $^{115}$Rh was produced at the University of Jyv\"askyl\"a in Finland by proton-induced fission of $^{238}$U. The isotopes were separated with the IGISOL on-line isotope separator and identified by $\gamma$-ray, x-ray, and $\beta$-ray emissions. ``The half-life of the $\beta$ decay of $^{115}$Rh was determined to be 0.99$\pm$0.05~s.'' This value corresponds to the presently adopted half-life of $^{115}$Rh.

\subsection{$^{116}$Rh}\vspace{0.0cm}

In ``A study of the low-energy transitions arising from the prompt de-excitation of fission fragments,'' published in 1970, Watson et al. reported the first observation of $^{116}$Rh \cite{1970Wat01}. Fission fragments from the spontaneous fission of $^{252}$Cf were measured in coincidence with X-rays and conversion electrons. In a table, the energy (49~keV) and half-life (0.82~ns) of an excited state were identified. These values are mentioned in the ENSDF database but the level has not been placed in the level scheme \cite{2008ENS01}.

\subsection{$^{117}$Rh}\vspace{0.0cm}

``Identification of the rare neutron-rich isotope $^{117}$Rh'' was published by Penttila et al. in 1991 reporting the observation of $^{117}$Rh \cite{1991Pen01}. $^{238}$U targets were irradiated with 23 MeV protons from the Louvain-La Neuve cyclotron facility. $^{117}$Rh was separated with the LISOL ion guide setup and identified by measuring $\gamma$- and $\beta$-ray spectra. ``A beta half-life of 0.44$\pm$0.04 s was measured for this nucleus from the decay of the beta-coincident K x rays of Pd. Three gamma rays of 34.6, 131.7, and 481.6 keV were found to be associated with the decay of $^{117}$Rh.'' The measured half-life corresponds to the presently adopted value for $^{117}$Rh.

\subsection{$^{118-121}$Rh}\vspace{0.0cm}

Bernas et al. discovered $^{118}$Rh, $^{119}$Rh, $^{120}$Rh, and $^{121}$Rh in 1994 at GSI, Germany, as reported in ``Projectile fission at relativistic velocities: A novel and powerful source of neutron-rich isotopes well suited for in-flight isotopic separation'' \cite{1994Ber01}. The isotopes were produced using projectile fission of $^{238}$U at 750 MeV/nucleon on a lead target. ``Forward emitted fragments from $^{80}$Zn up to $^{155}$Ce were analyzed with the Fragment Separator (FRS) and unambiguously identified by their energy-loss and time-of-flight.'' The experiment yielded 313, 82, 13, and 3 individual counts of $^{118}$Rh, $^{119}$Rh, $^{120}$Rh, and $^{121}$Rh, respectively.

\subsection{$^{122}$Rh}\vspace{0.0cm}

$^{122}$Rh was discovered by Bernas et al. in 1997, as reported in ``Discovery and cross-section measurement of 58 new fission products in projectile-fission of 750$\cdot$AMeV $^{238}$U'' \cite{1997Ber01}. The experiment was performed using projectile fission of $^{238}$U at 750~MeV/nucleon on a beryllium target at GSI in Germany. ``Fission fragments were separated using the fragment separator FRS tuned in an achromatic mode and identified by event-by-event measurements of $\Delta$E-B$\rho$-ToF and trajectory.'' During the experiment, five counts for $^{122}$Rh were recorded.

\subsection{$^{123-126}$Rh}\vspace{0.0cm}

The discovery of $^{123}$Rh, $^{124}$Rh, $^{125}$Rh, and $^{126}$Rh was reported in the 2010 article ``Identification of 45 new neutron-rich isotopes produced by in-flight fission of a $^{238}$U beam at 345 MeV/nucleon,'' by Ohnishi et al. \cite{2010Ohn01}. The experiment was performed at the RI Beam Factory at RIKEN, where the new isotopes were created by in-flight fission of a 345 MeV/nucleon $^{238}$U beam on beryllium and lead targets. The isotopes were separated and identified with the BigRIPS superconducting in-flight separator. The results for the new isotopes discovered in this study were summarized in a table. 933 counts for $^{123}$Rh, 94 counts for $^{124}$Rh, 13 counts for $^{125}$Rh, and 1 count for $^{126}$Rh were recorded.

\section{Summary}

The discoveries of the known rubidium, strontium, molybdenum, and rhodium isotopes have been compiled and the methods of their production discussed.
Compared to other elements in this mass region the discovery of the isotopes was fairly straightforward. Only a few ($^{79}$Rb, $^{78,79}$Sr, $^{88,89,104,105}$Mo, and $^{105}$Rh) isotopes were initially identified incorrectly. The half-lives of others ($^{86,88,91}$Rb, $^{91,92,93,96,97}$Sr, $^{91}$Mo and $^{97,98,99,106,107}$Rh) were reported first without mass assignments.

Curious was the simultaneous publication of four papers reporting the observation of $^{79}$Rb. The identification of $^{101}$Mo was especially difficult because of the almost identical half-lives of $^{101}$Mo and $^{101}$Tc. Maurer and Ramm published the solution simultaneously with Hahn and Strassmann, but Hahn and Strassmann gave Maurer and Ramm the credit for solving the puzzle. The discovery of $^{104}$Rh is generally credited to Pontecorvo, however, the paper by Pontecorvo misidentified the observed half-life as $^{105}$Rh.

\ack

This work was supported by the National Science Foundation under grants No. PHY06-06007 (NSCL) and PHY07-54541 (REU).

\bibliography{../isotope-discovery-references}

\newpage

\newpage

\TableExplanation

\bigskip
\renewcommand{\arraystretch}{1.0}

\section{Table 1.\label{tbl1te} Discovery of rubidium, strontium, molybdenum, and rhodium isotopes }
\begin{tabular*}{0.95\textwidth}{@{}@{\extracolsep{\fill}}lp{5.5in}@{}}
\multicolumn{2}{p{0.95\textwidth}}{ }\\

Isotope & Rubidium, strontium, molybdenum, and rhodium isotope \\
First author & First author of refereed publication \\
Journal & Journal of publication \\
Ref. & Reference \\
Method & Production method used in the discovery: \\

  & FE: fusion evaporation \\
  & NC: Neutron capture reactions \\
  & LP: light-particle reactions (including neutrons) \\
  & MS: mass spectroscopy \\
  & SF: spontaneous fission \\
  & NF: neutron induced fission \\
  & CPF: charged-particle induced fission \\
  & SP: spallation reactions \\
  & PF: projectile fragmentation of fission \\

Laboratory & Laboratory where the experiment was performed\\
Country & Country of laboratory\\
Year & Year of discovery \\
\end{tabular*}
\label{tableI}

\datatables 



\setlength{\LTleft}{0pt}
\setlength{\LTright}{0pt}


\setlength{\tabcolsep}{0.5\tabcolsep}

\renewcommand{\arraystretch}{1.0}

\footnotesize 

\begin{longtable}{@{\extracolsep\fill}llllllll@{}}
\caption{Discovery of Rubidium, Strontium, Molybdenum, and Rhodium Isotopes. See page\ \pageref{tbl1te} for Explanation of Tables}
Isotope & First Author & Journal & Ref. & Method & Laboratory & Country & Year\\
\hline\\
\endfirsthead\\
\caption[]{(continued)}
Isotope & First author & Journal & Ref. & Method & Laboratory & Country & Year\\
\hline\\
\endhead
$^{73}$Rb & J.C. Batchelder & Phys. Rev. C &\cite{1993Bat02}& FE & Berkeley & USA &1993 \\
$^{74}$Rb & J.M. D'Auria & Phys. Lett. B &\cite{1977DAu01}& SP & CERN & Switzerland &1977 \\
$^{75}$Rb & H.L. Ravn & J. Inorg. Nucl. Chem. &\cite{1975Rav01}& SP & CERN & Switzerland &1975 \\
$^{76}$Rb & J. Chaumont & Phys. Lett. B &\cite{1969Cha01}& SP & CERN & Switzerland &1969 \\
$^{77}$Rb & R. Arlt & Nucl. Instrum. Meth. &\cite{1972Arl02}& SP & Dubna & Russia &1972 \\
$^{78}$Rb & C. J. Toeset & Radiochim. Acta &\cite{1968Toe01}& LP & Amsterdam & Netherlands &1968 \\
$^{79}$Rb & R. Chaminade & Nucl. Phys. &\cite{1957Cha02}& FE & Saclay & France &1957 \\
$^{80}$Rb & R.W. Hoff & J. Inorg. Nucl. Chem. &\cite{1961Hof01}& FE & Berkeley & USA &1961 \\
$^{81}$Rb & F. L. Reynolds & Phys. Rev. &\cite{1949Rey01}& FE & Berkeley & USA &1949 \\
$^{82}$Rb & F. L. Reynolds & Phys. Rev. &\cite{1949Rey01}& FE & Berkeley & USA &1949 \\
$^{83}$Rb & D. G. Karraker & Phys. Rev. &\cite{1950Kar01}& LP & Berkeley & USA &1950 \\
$^{84}$Rb & W. C. Barber & Phys. Rev. &\cite{1947Bar01}& LP & Berkeley & USA &1947 \\
$^{85}$Rb & F.W. Aston & Nature &\cite{1921Ast03}& MS & Cambridge & UK &1921 \\
$^{86}$Rb & A. C. Helmholz& Phys. Rev. &\cite{1941Hel01}& LP & Berkeley & USA &1941 \\
$^{87}$Rb & F.W. Aston & Nature &\cite{1921Ast03}& MS & Cambridge & UK &1921 \\
$^{88}$Rb & F. A. Heyn & Nature &\cite{1939Hey01}& NF & Eindhoven & Netherlands &1939 \\
$^{89}$Rb & G.N. Glasoe & Phys. Rev. &\cite{1940Gla02}& NF & Columbia & USA &1940 \\
$^{90}$Rb & O. Kofoed & Phys. Rev.&\cite{1951Kof01}& NF & Copenhagen & Denmark &1951 \\
$^{91}$Rb & O. Kofoed & Phys. Rev.&\cite{1951Kof01}& NF & Copenhagen & Denmark &1951 \\
$^{92}$Rb & K. Fritze & Can. J. Phys. &\cite{1960Fri01}& NF & McMaster & Canada &1960 \\
$^{93}$Rb & K. Fritze & Can. J. Phys. &\cite{1960Fri01}& NF & McMaster & Canada &1960 \\
$^{94}$Rb & K. Fritze & Can. J. Phys. &\cite{1961Fri01}& NF & McMaster & Canada &1961 \\
$^{95}$Rb & I. Amarel & Phys. Lett. B &\cite{1967Ama01}& CPF & Orsay & France &1967 \\
$^{96}$Rb & I. Amarel & Phys. Lett. B &\cite{1967Ama01}& CPF & Orsay & France &1967 \\
$^{97}$Rb & I. Amarel & J. Inorg. Nucl. Chem. &\cite{1969Ama01}& CPF & Orsay & France &1969 \\
$^{98}$Rb & B.L. Tracy & Phys. Lett. B &\cite{1971Tra01}& CPF & Grenoble & France &1971 \\
$^{99}$Rb & B.L. Tracy & Phys. Lett. B &\cite{1971Tra01}& CPF & Grenoble & France &1971 \\
$^{100}$Rb & E. Koglin & Z. Phys. A &\cite{1978Kog01}& NF & Grenoble & France &1978 \\
$^{101}$Rb & K. Balog & Z. Phys. A &\cite{1992Bal01}& SP & CERN & Switzerland &1992 \\
$^{102}$Rb & G. Lhersonneau & Z. Phys. A &\cite{1995Lhe01}& SP & CERN & Switzerland &1995 \\
$^{103}$Rb & T. Ohnishi & J. Phys. Soc. Japan &\cite{2010Ohn01}& PF & RIKEN & Japan &2010 \\
 & & & & & &  \\
 & & & & & &  \\
$^{73}$Sr & J.C. Batchelder & Phys. Rev. C &\cite{1993Bat02}& FE & Berkeley & USA &1993 \\
$^{74}$Sr & B. Blank & Phys. Rev. Lett. &\cite{1995Bla01} & PF & GANIL & France &1995 \\
$^{75}$Sr & M.F. Mohar & Phys. Rev. Lett. &\cite{1991Moh01}& PF & Michigan State & USA &1991 \\
$^{76}$Sr & C.J. Lister & Phys. Rev. C &\cite{1990Lis01}& FE & Daresbury & UK &1990 \\
$^{77}$Sr & J.C. Hardy & Phys. Lett. B &\cite{1976Har01}& FE & Chalk River & Canada &1976 \\
$^{78}$Sr & C.J. Lister & Phys. Rev. Lett. &\cite{1982Lis01}& FE & Brookhaven & USA &1982 \\
$^{79}$Sr & I.M. Ladenbauer-Bellis & Can. J. Phys. &\cite{1972Lad01}& FE & Yale & USA &1972 \\
$^{80}$Sr & R.W. Hoff & J. Inorg. Nucl. Chem. &\cite{1961Hof01}& FE & Berkeley & USA &1961 \\
$^{81}$Sr & S.V. Castner & Phys. Rev. &\cite{1952Cas01}& LP & Berkeley & USA &1952 \\
$^{82}$Sr & S.V. Castner & Phys. Rev. &\cite{1952Cas01}& LP & Berkeley & USA &1952 \\
$^{83}$Sr & S.V. Castner & Phys. Rev. &\cite{1952Cas01}& LP & Berkeley & USA &1952 \\
$^{84}$Sr & J.P. Blewitt & Phys. Rev. &\cite{1936Ble01} & MS & Princeton & USA &1936 \\
$^{85}$Sr & L.A. DuBridge & Phys. Rev. &\cite{1940DuB01}& LP & Rochester & USA &1940 \\
$^{86}$Sr & F.W. Aston & Nature &\cite{1924Ast04}& MS & Cambridge & UK &1931 \\
$^{87}$Sr & F.W. Aston & Nature &\cite{1931Ast01}& MS & Cambridge & UK &1931 \\
$^{88}$Sr & F.W. Aston & Nature &\cite{1923Ast01}& MS & Cambridge & UK &1923 \\
$^{89}$Sr & D.W. Stewart & Phys. Rev. &\cite{1937Ste01}& LP & Michigan & USA &1937 \\
$^{90}$Sr & W.E. Grummitt & Nature &\cite{1948Gru01}& NF & Chalk River & Canada &1948 \\
$^{91}$Sr & W. Seelmann-Eggebert & Naturwiss. &\cite{1943See02}& LP & Berln& Germany &1943 \\
$^{92}$Sr & G. Herrmann & Z. Naturforsch. &\cite{1956Her01}& NF & Mainz & Germany &1956 \\
$^{93}$Sr & J.D. Knight & J. Inorg. Nucl. Chem. &\cite{1959Kni01}& NF & Los Alamos & USA &1959 \\
$^{94}$Sr & J.D. Knight & J. Inorg. Nucl. Chem. &\cite{1959Kni01}& NF & Los Alamos & USA &1959 \\
$^{95}$Sr & K. Fritze & Can. J. Phys. &\cite{1961Fri01}& NF & McMaster & Canada &1961 \\
$^{96}$Sr & M.-I. Macias-Maruqes & J. Phys. (Paris) &\cite{1971Mac01}& CPF & Orsay & France &1971 \\
$^{97}$Sr & F.K. Wohn & Phys. Rev. C &\cite{1978Woh01}& NF & Grenoble & France &1978 \\
$^{98}$Sr & B.L. Tracy & Phys. Lett. B &\cite{1971Tra01}& CPF & Grenoble & France &1971 \\
$^{99}$Sr & M. Asghar & Nucl. Phys. A &\cite{1975Asg01}& NF & Grenoble & France &1975 \\
$^{100}$Sr & E. Koglin & Z. Phys. A &\cite{1978Kog01}& NF & Grenoble & France &1978 \\
$^{101}$Sr & F.K. Wohn & Phys. Rev. Lett. &\cite{1983Woh01}& NF & Brookhaven & USA &1983 \\
$^{102}$Sr & J.C. Hill & Phys. Rev. C &\cite{1986Hil01}& NF & Brookhaven & USA &1986 \\
$^{103}$Sr & M. Bernas & Phys. Lett. B &\cite{1997Ber01}& PF & Darmstadt & Germany &1997 \\
$^{104}$Sr & M. Bernas & Phys. Lett. B &\cite{1997Ber01}& PF & Darmstadt & Germany &1997 \\
$^{105}$Sr & M. Bernas & Phys. Lett. B &\cite{1997Ber01}& PF & Darmstadt & Germany &1997 \\
$^{106}$Sr & T. Ohnishi & J. Phys. Soc. Japan &\cite{2010Ohn01}& PF & RIKEN & Japan &2010 \\
$^{107}$Sr & T. Ohnishi & J. Phys. Soc. Japan &\cite{2010Ohn01}& PF & RIKEN & Japan &2010 \\
 & & & & & &  \\
 & & & & & &  \\
$^{83}$Mo & Z. Janas & Phys. Rev. Lett. &\cite{1999Jan01}& PF & GANIL & France &1999 \\
$^{84}$Mo & W. Gelletly & Phys. Lett. B &\cite{1991Gel01}& FE & Daresbury & UK &1991 \\
$^{85}$Mo & S.J. Yennello & Phys. Rev. C &\cite{1992Yen01}& PF & Michigan State & USA &1992 \\
$^{86}$Mo & C.J. Gross & Phys. Rev. C &\cite{1991Gro01}& FE & Daresbury & UK &1991 \\
$^{87}$Mo & G. Korschinek & Z. Phys. A &\cite{1977Kor01}& FE & Munich & Germany &1977 \\
$^{88}$Mo & T.A. Doron & Nucl. Phys. A &\cite{1971Dor01}& FE & Rochester & USA &1971 \\
$^{89}$Mo & R.C. Pardo & Phys. Rev. C &\cite{1980Par01}& LP & Michigan State & USA &1980 \\
$^{90}$Mo & R.M. Diamond & Phys. Rev. &\cite{1953Dia01}& LP & Harvard & USA &1953 \\
$^{91}$Mo & R. Sagane & Phys. Rev. &\cite{1938Sag02}& LP & Tokyo & Japan &1948 \\
$^{92}$Mo & F.W. Aston & Nature &\cite{1930Ast04}& MS & Cambridge & UK &1930 \\
$^{93}$Mo & M.L. Wiedenbeck & Phys. Rev. &\cite{1946Wie01}& LP & Michigan & USA &1946 \\
$^{94}$Mo & F.W. Aston & Nature &\cite{1930Ast04}& MS & Cambridge & UK &1930 \\
$^{95}$Mo & F.W. Aston & Nature &\cite{1930Ast04}& MS & Cambridge & UK &1930 \\
$^{96}$Mo & F.W. Aston & Nature &\cite{1930Ast04}& MS & Cambridge & UK &1930 \\
$^{97}$Mo & F.W. Aston & Nature &\cite{1930Ast04}& MS & Cambridge & UK &1930 \\
$^{98}$Mo & F.W. Aston & Nature &\cite{1930Ast04}& MS & Cambridge & UK &1930 \\
$^{99}$Mo & R. Sagane & Phys. Rev. &\cite{1938Sag02}& LP & Tokyo & Japan &1948 \\
$^{100}$Mo & F.W. Aston & Nature &\cite{1930Ast04}& MS & Cambridge & UK &1930 \\
$^{101}$Mo & W. Maurer & Naturwiss. &\cite{1941Mau01}& LP & Berlin & Germany &1941 \\
$^{102}$Mo & D.R. Wiles & Phys. Rev. &\cite{1954Wil01}& NF & MIT & USA &1954 \\
$^{103}$Mo & P. Kienle & Radiochim. Acta &\cite{1963Kie01}& NF & Munich & Germany &1963 \\
$^{104}$Mo & P. Kienle & Naturwiss. &\cite{1962Kie01}& NF & Munich & Germany &1962 \\
$^{105}$Mo & P. Kienle & Naturwiss. &\cite{1962Kie02}& NF & Munich & Germany &1962 \\
$^{106}$Mo & J.D. Hastings & Radiochim. Acta &\cite{1969Has01}& NF & Oak Ridge & USA &1969 \\
$^{107}$Mo & N. Trautmann & Radiochim. Acta &\cite{1972Tra01}& NF & Mainz & Germany &1972 \\
$^{108}$Mo & N. Trautmann & Radiochim. Acta &\cite{1972Tra01}& NF & Mainz & Germany &1972 \\
$^{109}$Mo & J. \"Ayst\"o& Phys. Rev. Lett. &\cite{1992Ays01}& CPF & Jyvaskyla & Finland &1992 \\
$^{110}$Mo & J. \"Ayst\"o& Phys. Rev. Lett. &\cite{1992Ays01}& CPF & Jyvaskyla & Finland &1992 \\
$^{111}$Mo & M. Bernas & Phys. Lett. B &\cite{1994Ber01}& PF & Darmstadt & Germany &1994 \\
$^{112}$Mo & M. Bernas & Phys. Lett. B &\cite{1994Ber01}& PF & Darmstadt & Germany &1994 \\
$^{113}$Mo & M. Bernas & Phys. Lett. B &\cite{1994Ber01}& PF & Darmstadt & Germany &1994 \\
$^{114}$Mo & M. Bernas & Phys. Lett. B &\cite{1997Ber01}& PF & Darmstadt & Germany &1997 \\
$^{115}$Mo & T. Ohnishi & J. Phys. Soc. Japan &\cite{2010Ohn01}& PF & RIKEN & Japan &2010 \\
$^{116}$Mo & T. Ohnishi & J. Phys. Soc. Japan &\cite{2010Ohn01}& PF & RIKEN & Japan &2010 \\
$^{117}$Mo & T. Ohnishi & J. Phys. Soc. Japan &\cite{2010Ohn01}& PF & RIKEN & Japan &2010 \\
 & & & & & &  \\
 & & & & & &  \\
$^{89}$Rh & K. Rykaczewski & Phys. Rev. C &\cite{1995Ryk01}& PF & GANIL & France &1995 \\
$^{90}$Rh & M. Hencheck & Phys. Rev. C &\cite{1994Hen01}& PF & Michigan State & USA &1994 \\
$^{91}$Rh & M. Hencheck & Phys. Rev. C &\cite{1994Hen01}& PF & Michigan State & USA &1994 \\
$^{92}$Rh & M. Hencheck & Phys. Rev. C &\cite{1994Hen01}& PF & Michigan State & USA &1994 \\
$^{93}$Rh & M. Hencheck & Phys. Rev. C &\cite{1994Hen01}& PF & Michigan State & USA &1994 \\
$^{94}$Rh & R.A. Zarifov & Bull. Acad. Sci. USSR &\cite{1979Zar01}& LP & Almaty & Kaszakhstan &1979 \\
$^{95}$Rh & A.H.W. Aten & Physica &\cite{1967Ate01}& LP & Amsterdam & Netherlands &1967 \\
$^{96}$Rh & A.H.W. Aten & Physica &\cite{1967Ate01}& LP & Amsterdam & Netherlands &1967 \\
$^{97}$Rh & A.H.W. Aten & Physica &\cite{1955Ate01}& LP & Amsterdam & Netherlands &1955 \\
$^{98}$Rh & A.H.W. Aten & Physica &\cite{1955Ate01}& LP & Amsterdam & Netherlands &1955 \\
$^{99}$Rh & C.L. Scoville & Phys. Rev. &\cite{1952Sco01}& LP & Ohio State & USA &1952 \\
$^{100}$Rh & M. Lindner & Phys. Rev. &\cite{1948Lin02}& SP & Berkeley & USA &1948 \\
$^{101}$Rh & M. Lindner & Phys. Rev. &\cite{1948Lin02}& SP & Berkeley & USA &1948 \\
$^{102}$Rh & O. Minakawa & Phys. Rev. &\cite{1941Min01}& LP & Tokyo & Japan &1941 \\
$^{103}$Rh & F.W. Aston & Nature &\cite{1934Ast02}& MS & Cambridge & UK &1934 \\
$^{104}$Rh & E.C. Crittenden & Phys. Rev. &\cite{1939Cri01}& NC & Cornell & USA &1939 \\
$^{105}$Rh & E. Bohr & Arkiv Mat. Astron. Fysik&\cite{1945Boh01}& LP & Stockholm & Sweden &1945 \\
$^{106}$Rh & W. C. Peacock & Phys. Rev. &\cite{1947Pea01}& NF & Oak Ridge & USA &1947 \\
$^{107}$Rh & L.E. Glendenin & Nat. Nucl. Ener. Ser. &\cite{1951Gle01}& NF & Oak Ridge & USA &1951 \\
$^{108}$Rh & G.B. Baro & Z. Naturforsch. &\cite{1955Bar01}& NF & Buenos Aires & Argentina &1955 \\
$^{109}$Rh & F. F. Hopkins & Phys. Rev. C &\cite{1972Hop01}& SF & Austin & USA &1972 \\
$^{110}$Rh & M. Karras & Phys. Lett. &\cite{1963Kar01}& LP & Arkansas & USA &1963 \\
$^{111}$Rh & G. Franz & Inorg. Nucl. Chem. Lett. &\cite{1975Fra02}& NF & Mainz & Germany &1975 \\
$^{112}$Rh & F. F. Hopkins & Phys. Rev. C &\cite{1972Hop01}& SF & Austin & USA &1972 \\
$^{113}$Rh & F. F. Hopkins & Phys. Rev. C &\cite{1971Hop01}& SF & Austin & USA &1971 \\
$^{114}$Rh & J. \"Ayst\"o & Nucl. Phys. A &\cite{1988Ays02}& CPF & Jyvaskyla & Finland &1988 \\
$^{115}$Rh & J. \"Ayst\"o & Phys. Lett. B &\cite{1988Ays01}& CPF & Jyvaskyla & Finland &1988 \\
$^{116}$Rh & R.L. Watson & Nucl. Phys. A &\cite{1970Wat01}& SF & Berkeley & USA &1970 \\
$^{117}$Rh & H. Penttil\"a & Phys. Rev. C &\cite{1991Pen01}& CPF & Louvain-la-Neuve & Belgium &1991 \\
$^{118}$Rh & M. Bernas & Phys. Lett. B &\cite{1994Ber01}& PF & Darmstadt & Germany &1994 \\
$^{119}$Rh & M. Bernas & Phys. Lett. B &\cite{1994Ber01}& PF & Darmstadt & Germany &1994 \\
$^{120}$Rh & M. Bernas & Phys. Lett. B &\cite{1994Ber01}& PF & Darmstadt & Germany &1994 \\
$^{121}$Rh & M. Bernas & Phys. Lett. B &\cite{1994Ber01}& PF & Darmstadt & Germany &1994 \\
$^{122}$Rh & M. Bernas & Phys. Lett. B &\cite{1997Ber01}& PF & Darmstadt & Germany &1997 \\
$^{123}$Rh & T. Ohnishi & J. Phys. Soc. Japan &\cite{2010Ohn01}& PF & RIKEN & Japan &2010 \\
$^{124}$Rh & T. Ohnishi & J. Phys. Soc. Japan &\cite{2010Ohn01}& PF & RIKEN & Japan &2010 \\
$^{125}$Rh & T. Ohnishi & J. Phys. Soc. Japan &\cite{2010Ohn01}& PF & RIKEN & Japan &2010 \\
$^{126}$Rh & T. Ohnishi & J. Phys. Soc. Japan &\cite{2010Ohn01}& PF & RIKEN & Japan &2010 \\
 \\
\end{longtable}

\end{document}